\RequirePackage{fix-cm}
\documentclass[smallextended]{svjour3}       
\smartqed
\usepackage{graphicx}
\usepackage{todonotes}
\usepackage{subfig}
\usepackage{url}
\usepackage{tcolorbox}

\begin{document}
\title{Ethical Issues in Empirical Studies using Student Subjects: Re-visiting Practices and Perceptions}
\titlerunning{Ethical Issues in Empirical Studies using Student Subjects}        

\author{Grischa Liebel \and Shalini Chakraborty}

\institute{G. Liebel \at
              Reykjavik University\\ Menntavegur 1, 102 Reykjav\'{i}k, Iceland \\ ORCID: 0000-0002-3884-815X \\
              \email{grischal@ru.is}      
              \and
              S. Chakraborty \at
              Reykjavik University\\ Menntavegur 1, 102 Reykjav\'{i}k, Iceland \\
              \email{shalini19@ru.is}      
}

\date{Received: date / Accepted: date}

\maketitle

\begin{tcolorbox}
\textbf{Self-archiving note:}\\
This is a post-peer-review, pre-copyedit version of an article published in Empirical Software Engineering. The final authenticated version is available on Springer Link via \url{https://dx.doi.org/10.1007/s10664-021-09958-4}
\end{tcolorbox}

\begin{abstract}
\emph{Context:} Using student subjects in empirical studies has been discussed extensively from a methodological perspective in Software Engineering (SE), but there is a lack of similar discussion surrounding ethical aspects of doing so. As students are in a subordinate relationship to their instructors, such a discussion is needed.
\emph{Objective:} We aim to increase the understanding of practices and perceptions SE researchers have of ethical issues with student participation in empirical studies.
\emph{Method:} We conducted a systematic mapping study of 372 empirical SE studies involving students, following up with a survey answered by 100 SE researchers regarding their current practices and opinions regarding student participation.
\emph{Results:} 
The mapping study shows that the majority of studies does not report conditions regarding recruitment, voluntariness, compensation, and ethics approval. In contrast, the majority of survey participants supports reporting these conditions. The survey further reveals that less than half of the participants require ethics approval. Additionally, the majority of participants recruit their own students on a voluntary basis, and use informed consent with withdrawal options. There is disagreement among the participants whether course instructors should be involved in research studies and if should know who participates in a study.
\emph{Conclusions:}
It is a positive sign that mandatory participation is rare, and that informed consent and withdrawal options are standard. However, we see immediate need for action, as study conditions are under-reported, and as opinions on ethical practices differ widely. In particular, there is little regard in SE on the power relationship between instructors and students.
\end{abstract}

\section{Introduction}
Using student subjects in empirical studies is common in Software Engineering (SE).
In comparison to professionals, students are typically easier to recruit, more homogeneous in terms of skills and experience, and available in larger numbers \cite{falessi18}.
Furthermore, students may serve as a representative sample in many situations, e.g., if the target population is junior developers \cite{falessi18}.

While multiple studies exist in SE that discuss the validity of using student subjects in empirical research, e.g., \cite{falessi18,host00,runeson03,salman15,svahnberg08}, there is a lack of research investigating ethical considerations of student use.
Furthermore, early work in empirical SE shows little awareness and even disregard for ethical issues \cite{hall01}.
As students are under the control of their instructor, using students subjects raises several ethical considerations in addition to the issues already present in studies involving human subjects \cite{singer02,sieber01}.
Among others, it is necessary to ensure that students achieve their educational goals, and that empirical studies do not negatively affect learning outcomes by competing for limited time and resources \cite{carver10}.
Therefore, we investigate in this paper how student subjects are treated in SE research\footnote{Note that we target SE-focused research. Therefore, we do not specifically target SE education or computer science education research.}, and how this picture compares to several studies in the early 2000s.
We aim to answer the following research questions.
\begin{itemize}
    \item[RQ1] What are the conditions under which student subjects participate in empirical studies (in SE)?
    \item[RQ2] As a community, what do we believe are the ideal conditions under which student subjects participate in empirical studies (in SE)?
\end{itemize}
For the conditions, we focus on the research method applied, the voluntariness of participation, whether ethics approval is present, and the compensation of study participation.
To answer RQ1, we conducted a systematic mapping study, extracting and screening publications that use student subjects in top SE venues\footnote{The complete list of venues is found in Appendix \ref{app:venues}.}.
Based on the results of this study, we additionally conducted a survey with authors\footnote{To avoid confusion, throughout the paper we refer to the authors participating in our survey as ``participants'', while referring to students participating in a study as ``subjects''.} of the primary studies regarding their opinions on student subject use in SE to answer RQ2.

Our findings are that student subjects are typically used in controlled experiments.
The majority of studies recruit students on a voluntary basis, often as a part of a course.
Ethics approvals and compensation is only rarely mentioned in the primary studies.

The survey shows that only 43\%  of participants need ethics approval.
The majority uses their own students, and on a voluntary basis.
Students are compensated by 50\% of the participants.
Informed consent and withdrawal options are used by more than 90\% of participants.
Participants disagree on how acceptable different practices are, in particular with respect to the researcher-student relationship.

Finally, we find a substantial gap between the wish of our survey participants that study conditions should be reported, and the actual picture drawn by our systematic mapping study in which study conditions are only rarely reported.

With this paper, we hope to contribute to the discussion surrounding ethical use of student subjects in SE, and to increase the awareness of ethical standards and guidelines. \section{Related Work}

Several publications are of direct interest to this study, either as they describe general ethical principles in relation to SE, e.g., \cite{singer02,sieber01,andrews01,hall01}, or as they describe ethical principles or general guidelines in relation to student subjects, e.g. \cite{carver03,carver10,galster12,buse11,ko15}.

Singer and Vinson~\cite{singer02} list ethical considerations compiled from existing codes of ethics in other disciplines, e.g., other engineering disciplines.
The authors highlight informed consent, stating that, while ethicists ``do not fully agree on the necessary components'', it should contain elements such as voluntariness, the consent decision, and the right to withdraw from the study at any time.
Furthermore, they highlight that the power of a course instructor is problematic.
Even if informed consent is in place, students might fear reprisal if they do not participate.
Specifically, this is the case regardless of the instructor's intent, ``the ethical difficulty arises not from the professor’s intent but from her power.'' \cite{singer02}.
The authors suggest that students can remain anonymous to the instructor by using anonymous surveys/data collection, or the help of a graduate student to administer the data collection.
Finally, the authors clarify that consent is typically not required when there is no private data identifying the students in the raw data, i.e., if the students can expect to remain private.

As a more detailed follow-up publication, Vinson and Singer~\cite{vinson08} provide guidelines for ethical research involving humans in SE.
Specifically, the authors discuss four pillars that should be required in all such studies: use of informed consent, beneficence to subjects or reduction of harm, confidentiality of subjects and information they share, and scientific value of the study.
The authors further provide guidelines for ethics reviews.
Finally, they conclude stating that ethics in empirical studies needs to be considered specifically, and that researchers need to be trained to do so.

Andrews and Pradhan~\cite{andrews01} discuss ethical considerations in empirical SE.
Interestingly, they essentially do not mention student subjects as a specific area of concern.
As an exception, the authors mention that informed consent must be used, with ``full disclosure of all potentially adverse effects'' and option to withdraw at any time.

Davison et al.~\cite{davison01} summarise a panel discussion in the information systems community on research ethics.
The authors conclude that voluntary participation in empirical studies should be the norm, and that students should be informed how their data is handled.
Furthermore, the authors state that a code of practice is likely more successful than strict regulation, since it encourages a constructive dialogue in the research community.

Hall and Flynn~\cite{hall01} report on a survey conducted among SE researchers in the UK regarding attitudes towards ethical issues in empirical SE.
The authors state that there is only little attention towards ethical issues with human participation in SE, compared to traditional disciplines such as psychology, medicine, or law.
Their results show a worrying picture, with a lack of awareness and concern among participants for ethical issues.

Sieber \cite{sieber01} discusses specifically how to protect study subjects in empirical SE.
The author explains that risks (e.g., inconvenience, economic risk), context (e.g., students as subjects), and vulnerabilities (e.g., subordination of subjects) need to be considered when designing studies.
As in \cite{singer02}, the author discusses that pressure on students arises ``even if they are assured that participation [in a study] is voluntary.''~\cite{sieber01}.
Similarly, to lower this risk, the author suggests using graduate students to preserve anonymity of the subjects towards the instructor.

According to the different ethical concerns listed by Singer and Vinson~\cite{singer02}, Storey et al.~\cite{storey01} discuss ethical issues in relation to an empirical study conducted in their own course, after ethics approval was obtained.
The authors state that, while informed consent was obtained, they were unsure whether they did not inadvertently coerce students into participating, e.g., by offering course credits and unfavourable alternative assignments.
Similarly, they note that there might have been a too tight connection between instructors and researchers, thus putting additional stress on the participating students.
The authors conclude that many ethical issues arise from exposing students to tools that are untested, such as research prototypes.

Carver et al.~\cite{carver03} discuss ethical issues arising through the use of student subjects in empirical SE studies.
The authors contrast the students' ``rights to reach their educational goals'' with the costs caused by study participation.
Several ethical questions are raised, most of which serve as direct source for the hypotheses used in our study (see Section~\ref{sec_method}).
For instance, the authors ask whether it is ethically correct ``to base some of the final evaluation of a student on his or her performance in the empirical study''.

In a more recent publication by Carver et al.~\cite{carver10}, the ethical questions are complemented with requirements for empirical studies using student subjects.
As a general problem, the authors state that an empirical study needs to compete for ``scarce time, effort, and resources'' in university courses.
Especially relevant for our study are three requirements raised by the authors, namely that ethical issues must be addressed by the study design, that student should understand the value of empirical studies and how to conduct them, and that group projects should be included in empirical studies with student participation.
At the same time, several issues raised by \cite{singer02,sieber01,andrews01} are not addressed by the requirements.
For instance, the authors do not discuss that it can be problematic if the instructor is aware of which students participated in a study, as this might put stress on students or coerce them into participating.

Experiences with the guidelines by \cite{carver10} are discussed by Galster et al. \cite{galster12}.
Especially relevant to our study is the observation that it is helpful for the course instructor and the researcher conducting the study to be the same person, since it ``reduces problems with communicating the pedagogical value to students'' and since students ``felt less intimidated compared to a situation in which an external person conducts the study''.
This contrasts with statements from related work on ethics, e.g., \cite{singer02,sieber01}, that students feel pressure to participate when researcher and course instructor are identical.

Buse et al.~\cite{buse11} study publications at the CHI conference series with respect to ethics approval, following a similar method as we do in this study.
The authors find that only 14 out of 211 studies mention ethics approval.
Additionally, the authors find that ethics approval procedures are a perceived barrier in user studies.

Ko et al.~\cite{ko15} formulate guidelines for controlled experiments with human subjects in SE.
The authors state that informed consent should be used, but that ``it is likely that they [the authors] are required to obtain informed consent and that the study design itself must first be approved by an ethics committee''.
Similar to \cite{buse11}, the authors state that this approval process can be time-consuming.

Finally, recent work investigates ethics in SE in relation to emerging research areas, such as bots or autonomous systems, e.g., \cite{bowser15,buchanan11,aycock11}.
While this work is outside of our scope, the named publications re-iterate the basic principles named in \cite{vinson08}, i.e., informed consent, beneficence, confidentiality, and scientific value. 
    
In summary, work from the early 2000s reports that there is little awareness and concern for ethical issues in SE research \cite{hall01}.
Consequently, existing work summarises general ethical guidelines and procedures related to studies with human subjects \cite{singer02,sieber01,andrews01,hall01} and, in particular, with student subjects \cite{carver03,carver10,galster12,buse11,ko15}..
This body of work extensively references existing work in other disciplines.
In addition to this body of work, there is existing work that formulates SE-specific guidelines \cite{carver03,carver10}.
While these are often concerned with validity of the studies, there exists also ethical advice related to student subject use.
Interestingly, we find a disconnect between these two areas: Certain concerns mentioned in the more general work on ethics are seemingly ignored in SE-specific guides.
Furthermore, ethical issues remain even in cases where ethics approval is obtained \cite{storey01}.
Finally, we find that some advice is in direct conflict with general ethical advice, e.g., the supposed advantage of a course instructor being the same person as the researcher conducting the study \cite{galster12}.
This raises the question which practices, if any, are currently followed in the SE field, and what beliefs exist among SE researchers related to how ethical those practices are.

 \section{Research Method}
\label{sec_method}
The aim of this paper is to improve the understanding of ethical issues pertaining to the conditions under which they participate in empirical studies in SE.
We focus on the research method applied, the voluntariness of participation, whether ethics approval is present, and the compensation of study participation.
To address this aim, we formulate the following two RQs:
\begin{itemize}
    \item[RQ1] What are the conditions under which student subjects participate in empirical studies (in SE)?
    \item[RQ2] As a community, what do we believe are the ideal conditions under which student subjects participate in empirical studies (in SE)?
\end{itemize}
To answer RQ1, we conducted a literature study of empirical studies published in top SE venues.
While Systematic Literature Reviews (SLRs) have become common in SE research \cite{petersen08}, their purpose is to provide an in-depth review of the results and methodology of a number of papers \cite{petersen08}.
To answer our questions, we instead need to provide an overview of a large number of publications in the field of SE, without the need to review the results and methodology in depth.
Therefore, we instead chose to conduct a systematic mapping study as this is an appropriate way to provide this overview \cite{petersen08}, and as it allows to handle a larger number of primary studies \cite{petersen08}.

To answer RQ2, we conducted a survey among SE researchers publishing empirical studies with student participation.
We chose a survey design since surveys can allow for generalisation over a population of actors \cite{stol18}, in this case SE researchers experienced in conducting studies with students, and with an interest in ethics.

The designs of the two studies are explained in the following sub-sections.

\subsection{Systematic Mapping Study}
We break down RQ1 into a number of sub-research questions.
\begin{itemize}
\item[RQ1.1] In which top SE venues are empirical studies with student subjects published?
\item[RQ1.2] What are the most frequently applied research strategies in which student subjects are involved?
\item[RQ1.3] How many students are participating in the found studies?
\item[RQ1.4] To what extent is ethics approval reported in empirical studies using student subjects?
\item[RQ1.5] To what extent is participation of students in empirical studies voluntary?
\item[RQ1.6] How is the participation of students in empirical studies compensated?
\end{itemize}

Based on our research questions, we formulated the search string in accordance with \cite{kitchenham07}, breaking down our research questions into individual facets.
In our case, we consider all empirical studies (study design) from software engineering (population) that include students (context).
We did not find any synonyms to these keywords and, therefore, decided to use the broad search string
\emph{``software engineering'' AND ``empirical'' AND ``student''}.
We searched Scopus, IEEE Xplore and ACM Digital Library using the search string adapted to the format used in the respective database.
For IEEE Xplore and ACM Digital Library, we searched the full-text version of the paper.
For Scopus, full-text search is not available, so we searched in title, abstract and keywords instead.

We limited the search to papers published since 2010, since Carver et al.~\cite{carver10} published their updated checklist on student empirical studies in SE in that year.
Clearly, including earlier years would add valuable information and allow for a better analysis of trends.
However, we decided that the Carver et al.~\cite{carver10} study marks a logical point in time to restrict the search, and at the same time limits the effort since the number of papers already grew rather large in comparison to many existing studies.
Similarly, we chose to only include papers from high-quality SE venues in our search, since we wanted to obtain a picture of SE research that would be considered to be of high quality by scholars in the field.
As \emph{high quality} venues, we decided to use all SE conferences ranked A or A* in the most recent CORE conference ranking\footnote{\url{http://portal.core.edu.au/conf-ranks/}}, and the 14 SE journals included in the ranking by Robert Feldt\footnote{\url{http://www.robertfeldt.net/advice/se_venues/}}.
While this selection of venues is, to some extent, an arbitrary choice, we believe that slight differences in the in/exclusion of venues would not introduce a significant change in our results.
The entire list of venues is depicted in Appendix \ref{app:venues}.
Note that the three databases we searched index all of these venues to some extent.
However, certain titles are not completely included, or not full-text searchable.
For instance, JSEP, STVR, SPE, and IJSEKE were only searched on title, abstract and keywords, and for HICSS not all conference years are indexed in the used databases.

The search yielded 1284 papers, 1140 after duplicate removal.
For the remaining papers, we applied the following exclusion criteria on the title and abstract:
\begin{itemize}
\item Paper is less than 8 pages in length (short papers).
\item Title/abstract does not clearly indicate that the paper contains a primary empirical study.
\item From the title/abstract, it is evident that data was not collected from student subjects. This criterion also excludes secondary/tertiary studies such as systematic literature reviews.
\end{itemize}
Regarding the last criterion, we excluded studies that, e.g., made statements such as ``we conducted a controlled experiment with 50 professional software developers'', with no indication of further data collection.
If unclear, we would include the paper in the next step.
It can be argued that papers with less than 8 pages in length should also be included, as they might provide valuable information.
However, we decided against that, as many venues consider those short papers or other special-track papers, such as new ideas and visions, with a specific focus.
Due to the length and their specific focus, it is less likely that such papers report study conditions.
For instance, a new ideas and vision paper will likely spend more time elaborating on the novelty of the idea, rather than on a performed or planned study design.

Based on these exclusion criteria, we excluded 652 papers, leaving us with 488 papers.
For these 488 papers, we obtained the full-text version of each paper.
Applying the exclusion criteria of title/abstract on the full-text versions, we excluded another 112 papers.
Finally, 4 papers were behind a paywall and not available to us.
This left us with 372 papers for analysis.

Based on targeted reading of the introduction and method sections, as well as keyword search, we extracted the information relevant to our review, i.e., study type, venue, number of student subjects, voluntariness of participation, subject compensation, and status of ethics approval.
The detailed extraction procedures are listed in Appendix~\ref{app:dataExtraction}, including the keywords we searched for.
In contrast to the recommendations by Petersen et al.~\cite{petersen08}, we extracted this information from the full-text publication.
We visualised the resulting information using summary statistics and bubble plots.

Initially, we did not perform any reliability checks, as we deemed the exclusion and extraction process to be comparably objective.
However, in a first revision of this paper we added inter-rater and intra-rater reliability checks in a post-study fashion.
We aimed for a Cohen's Kappa of $\kappa \geq 0.61$ for all inter-rater reliability checks, and a $\kappa \geq 0.81$ for all intra-rater reliability checks, meaning substantial and perfect agreement according to Landis and Koch~\cite{landis77}.
To conduct the checks, the first and second author independently processed a random set of 10\% of the initial 1140 papers, applying the exclusion criteria on the title and abstracts.
Additionally, the first author re-did his exclusion approximately 20 months after the original process.
We obtained an inter-rater agreement of $\kappa \approx 0.722$ and an intra-rater agreement of $\kappa \approx 0.836$.
Given that we reached the desired values, we continued with the reliability checks for the full-text exclusion and extraction, without re-processing the original study set.

For the full-text exclusion and extraction, we again selected randomly 10\% of the papers included after the first stage and applied the extraction guidelines listed in Appendix~\ref{app:dataExtraction}.
To calculate the Cohen's Kappa scores, we first calculated whether there was sufficient agreement on potential exclusion of papers at this stage (which there was both for inter-rater and intra-rater agreement), then calculated the scores for the extracted values on voluntariness, compensation, ethics approval, and study type.
Since study type and compensation were taken verbatim from the paper, we checked in a first step (qualitatively, without a defined process) whether the values of the two raters could indeed be considered equivalent.
For inter-rater reliability, we recorded substantial agreement on voluntariness ($\kappa \approx 0.606$) and compensation ($\kappa \approx 0.937$).
However, we had lower reliability on ethics approval ($\kappa \approx 0.552$) and study type ($\kappa \approx 0.378$).
Analysing the disagreements showed that the second author had rated the use of informed consent and withdrawal as a 'yes' on ethics approval, even if there was no explicit approval.
Similarly, she had described both families and replications of controlled experiments simply as controlled experiments, while the first author had gone into more depth.
After discussing those issues, the second author re-did her extraction (without seeing the first author's results), yielding substantial agreement also for ethics approval ($\kappa \approx 0.787$) and study type ($\kappa \approx 0.713$).
The intra-rater agreement showed perfect agreement on ethics approval ($1.0$), and compensation ($.939$).
Only substantial agreement was achieved for voluntariness ($0.778$) and study type ($0.706$). 
For voluntariness, the lower agreement was due to the first author being inconsistent in the original extraction for the distinction between no information on voluntariness ('NA') and being part of a course, with no further information ('part of a course').
Therefore, we decided to extract the voluntariness again for all full-text papers that were originally rated 'NA', leading to 29 changes.
For study type, the lower agreement was due to primary studies describing their methodology in multiple ways.
For instance, a study could at some point in the paper be described as an ``evaluation study'', and later on as a ``controlled experiment'', leading to two different extractions.
We argue that, for study type, a substantial agreement is sufficient given that the field uses inconsistent terminology when it comes to methodology \cite{stol18}.

The mapping study dataset is available at \cite{datasetThis}.

\subsection{Researcher Survey}
The overview obtained from the systematic map provides us with information on empirical studies involving students published in high-quality SE venues.
However, while this picture is likely to be representative for high-quality SE publications, it might not reflect well the current perceptions of SE researchers regarding student subjects in empirical studies.
For instance, publication bias or compromises during study design, data collection and analysis might lead to a skewed picture.

Hence, we decided to follow up our systematic map with a survey among SE researchers that conduct empirical studies using students\footnote{For this survey, we did not need to obtain ethics approval at our institution. Participation was voluntary and no incentive was provided. The first survey page as well as the invitation email stated which data was collected and how the obtained data would be used.}.
To design the survey, we followed the guidelines by Kitchenham and Pfleeger~\cite{kitchenham08,pfleeger01}.

The \textbf{objective} of this descriptive follow-up survey was to understand whether or not the perceptions of SE researchers regarding student participation in empirical studies are in line with published accounts.
Based on the mapping study results, on the updated guidelines released by Carver et al.~\cite{carver10}, and our experiences, we formulated a number of hypotheses that guided us in the design of our survey.
We assume that the mapping study results are reflecting researcher's beliefs about study participation, that the guidelines by Carver et al.~\cite{carver10} are being applied, and that our experiences are representative for SE research.
The hypotheses, together with their sources, are depicted below.
\begin{table}
\caption{Hypotheses and their Sources}
\label{tab:hypotheses}
\begin{tabular}{|p{0.05\textwidth}|p{0.6\textwidth}|p{0.3\textwidth}|}
\hline
ID & Hypothesis & Source\\
\hline
$H_1$ & The majority of participants are not required to obtain ethics approval
        & (Mapping study, \cite{buse11}) \\
\hline
$H_2$ & The majority of participants recruit their own students for empirical studies. & (Experience, no mapping study data) \\
\hline
$H_3$ & The majority of participants perform voluntary studies, as a part of courses. & (Mapping study) \\
\hline
$H_4$ & The majority of participants do not offer compensation to their subjects. & (based on inconclusive mapping study data) \\
\hline
$H_5$ & The majority of participants use informed consent, including the option to withdraw voluntarily. & (\cite{carver10}, no mapping study data) \\
\hline
$H_6$ & The majority of participants agree that studies should relate to course learning outcomes or project work. & (sound ``reasonable'', \cite{carver10}) \\
\hline
$H_7$ & The majority of participants agree that informed consent and withdrawal should be offered. & (\cite{carver10}, no mapping study data) \\
\hline
$H_8$ & The majority of participants disagree that information on ethics approval, voluntariness, compensation and informed consent should be included in publications. & (mapping study) \\
\hline
\end{tabular}
\end{table}

We designed the survey as a \textbf{cross-sectional, self-administered questionnaire}, i.e., participants fill in the questionnaire through an online tool at a single point in time.
The survey questions were formulated based on the scope and the findings of the mapping study, related work on ethical issues in student subject studies, e.g., \cite{carver03,carver10,galster12}, and the hypotheses listed in Table~\ref{tab:hypotheses}.
The questionnaire followed an hourglass format, with general demographic questions at the beginning, increasingly detailed questions regarding the participants' opinions and past studies, and open questions/comment fields to end the survey.
Demographic questions were mainly open, to allow for different academic systems, e.g., in position descriptions.
For questions regarding past study conditions and opinions on student recruiting and reporting, we used closed questions with free-text clarification options.
We tried to design the questions following best practice, e.g., avoiding jargon, leading questions, or Yes/No questions (unless appropriate).
We did not try to measure concepts that are hard to map to single questions and would require summated rating scales \cite{kitchenham08} or similar.
The survey questionnaire is listed in Appendix~\ref{app:questionnaire}, and can be found in the dataset~\cite{datasetThis}.

The survey was instrumented using the online service \emph{SoSciSurvey}\footnote{https://soscisurvey.de}.
We included an introduction page with purpose statement, a declaration that the participation is voluntary and anonymous, and a statement regarding the use of the obtained data.
The length of the survey was designed in a way that it should not take more than 15 minutes, unless extensive free-text answers were given.
From the meta data, we know that only 3 participants exceeded this time, and one participant confirmed in an email that the length had been estimated well.

To \textbf{pilot the survey}, we sent it to two researchers, one in SE and one in higher education.
Furthermore, the latter researcher is a native English speaker.
The researchers answered the survey, and reviewed the questions in terms of content and form.
We did not perform a test-retest reliability assessment.

We followed a \textbf{purposeful sampling strategy}, selecting the first two authors from each publication included in the systematic mapping study.
Thus, we characterise the \textbf{population} as SE researchers experienced in conducting studies with student subjects. 
This does not allow us to answer RQ2 for SE researchers in general, but is likely to yield a higher response rate as we can expect the sample to be more interested in our research questions.

We selected the first two authors, since those are in many cases the ones most familiar with the published studies (due to the dominant practice in SE to order authors by contribution).
We then merged duplicate email addresses (in cases where one author appeared on more than one publication) and multiple email addresses for the same person (where we could identify them).
This led to a list of 504 recipients, who we contacted with an invitation for our survey.
For the 97 of those recipients who were not reachable, we found alternative email addresses in 48 cases.
For the remainder, we proceeded with the next author in the publication.
We repeated this step until we either obtained two valid email address per paper, or there were no further authors.
This led to a total of 474 delivered mail invitations.

We sent one reminder 10 days before the end of the survey administration period.
In total, we received 100 replies, corresponding to a 21.1\% response rate.

To analyse the survey data, we compared the visual representation of the data using bar graphs with the hypotheses.
Additionally, we used \textit{paradigmatic corroboration}~\cite{saldana15}.
That is, we coded the qualitative data obtained from free-text answers using descriptive coding \cite{saldana15}, i.e., assigning codes describing the discussed topic(s) similar to ``hash tags''.
We then checked whether or not they corroborate the quantitative survey data.
This coding process was not validated.

The survey dataset together with the complete instrument is available at \cite{datasetThis}.
Note that free-text answers have been anonymised where necessary, and are only shown in aggregate form to avoid individuals being identifiable based on their answers.
In particular, we anonymised mentions of countries with only one participant, and removed meta data (such as time spent on the survey).

\subsection{Threats to Validity}
In the following, we discuss the threats to the validity of the findings presented in this paper.
We structure these threats according to construct, internal, and external validity.
Additionally, we discuss the reliability of the studies.

\subsubsection{Construct Validity}
Construct validity reflects to what extent the measures represent the construct investigated.
In our study, we are investigating several aspects of ethics that could be misinterpreted by study participants.
For instance, the notion of a student \emph{subject} could have been misunderstood by participants to be limited only to controlled experiments, whereas students that are involved in a case study would not be considered subjects.
This is something we clarified in our survey after a participant notified us of the potential misunderstanding.

Furthermore, to increase validity, we piloted the survey with one colleague within SE, and another colleague outside of SE (with a role in university education) to review our survey instrument for clarity.

In the invitation for our survey, we presented invitees with preliminary results of our mapping study analysis.
Several of the invitees contacted us with feedback and/or additional impressions.
To these invitees, we sent an initial draft of this paper as a form of \textit{member checking}.
We decided not to ask for additional input from the other invitees, to limit the number of emails to those who might not be interested in the study.

\subsubsection{Internal Validity}

Internal validity reflects to what extend causal relationships are closely examined and other, unknown factors might impact the findings.

As a means to avoid oversimplification, we applied paradigmatic corroboration \cite{saldana15} during the survey analysis.
That is, we examined whether the qualitative, free-text answers corroborated the quantitative scores assigned to different answers.
This avoided, in some cases, a misinterpretation of the answers.
For instance, participants would assign a neutral score to a question, but then explain the free-text answers that they did not fully understand the question, or that it did not apply.

As a part of the results of the mapping study, we display the number of included papers per venue, normalised by the total amount of papers published at each venue during the time.
To do so, we manually extracted the number of published papers for each venue, relying on database searches (for journals) and information in conference proceeding preambles.
There is a minor threat that these numbers are not entirely correct, e.g., as database indices might be incomplete, or as conference proceedings might list numbers for multiple tracks instead of only the tracks we searched.

Since we screened for published papers, there is the threat that the results would differ when including grey literature, or rejected papers.
Intuitively, those papers could be less systematic in their practices and how they report them.
However, we might exclude papers which are under-represented due to community trends or bias.
For instance, qualitative studies are under-represented in SE \cite{storey19}.
This raises the potential threat of a systematic error.
Since we do not see a feasible way of systematic sampling those types of papers as well, we have to accept this threat to validity.

\subsubsection{External Validity}

External validity is reflecting to what extent findings can be generalised beyond the concrete sample.
In our mapping study, we screened publications from top SE venues.
We expect that the data obtained from this study is representative of high-quality SE publications.
Since quality criteria change over time, and vary between different venues, we do not expect to find similar results in earlier publications or in those published in venues that have more relaxed acceptance criteria.
In particular, we expect studies with less rigour (in terms of recruitment and sampling) and lower number of subjects per study.
In terms of study type, we would expect a larger variance in study type in other venues, since many of the top SE venues have been subject to criticism for potentially favouring quantitative studies.

Some of the venues we included in the mapping study are not accessible through full-text search.
The extent of this bias can be approximated.
Looking at ACM DL and IEEEXplore, the databases that allowed for full-text search, we observe that approximately 17.5\% of papers (17.13\% in ACM DL, 18.22\% in IEEEXplore) contain 'student' in the abstract/title/keywords.
The remainder of papers only include the term in the full-text search.
Of the papers that did not include the keyword in title/abs/keywords, 19.40\% were ultimately included at the full-text stage.
This is in contrast to the much larger 62.20\% of papers included where the keyword was found in title/abs/keywords.
Directly translating these numbers to the remaining database (Scopus) leads to an upper bound of papers we might have missed.
In Scopus, we found 702 papers where student was in title/abs/keywords. 
Assuming the same ratio as the IEEE and ACM databases would lead to another 3309 papers if Scopus would allow full-text search.
Of these, we would then include 19.40\%, meaning 642 additional full-text papers.
However, this is a very conservative estimate since Scopus overlaps to a large extent with the indexed venues of ACM DL and IEEE Xplore.
Excluding from the Scopus search results those papers published in venues that are at least partially indexed by ACM DL or IEEE Xplore, and re-doing the calculation, we end up with a lower bound of around 100-300 papers that might have been missed through the lack of full-text search.
Specifically, venues we are aware that are neither in ACM DL nor IEEEXplore are the Wiley Journals (Journal of Software: Evolution and Process, Software: Practice \& Experience, Software Testing, Verification and Reliability), World Scientific's Int. Journal of Software Engineering and Knowledge Engineering, and to a large extent HICSS conference.
While this bias is rather large, i.e., between 25\% and 172.58\%, we would expect the missing/incomplete venues not to report studies that differ substantially from the extracted ones.

Our survey sample was drawn from the population of SE researchers familiar with conducting and publishing empirical studies with student subjects.
We made this choice in order to allow for a higher response rate and an informed answer to our survey questions, as the sample can be considered more experienced with and interested in the study topic compared to the population of SE researchers in general.
Ideally, we would like to generalise our answer to RQ2 to SE researchers in general, but our sampling strategy does not allow this.
To allow for generalisation to SE research in general, a replication of the survey with different sampling strategy is needed.

Our survey response rate is 21.1\%, which is substantially higher than the typical 5\% stated by Lethbridge et al. \cite{lethbridge05}.
Considering that we have two contact emails for most papers, this means that we get on average approximately one answer for every second paper.
We deem this a reasonable response, but there might be a bias towards researchers most interested in ethical aspects.
We chose to only send a single reminder to not cause any annoyance, as survey invitations are becoming a nuisance in SE \cite{baltes16}.

A final threat to external validity is the exclusion of a number of email addresses in the survey.
In the original manuscript of this paper, we conducted the search in the 2019 version of the ACM Digital Library, which did not allow for full-text search.
In the first revision, we updated this search to include also full-text search, leading to the addition of 26 papers in the final full-text set.
Since we sampled the authors on the original data, this means we did not include the email addresses of these 26 papers, leading to 32 unique emails being omitted from the survey.
Assuming the same response rate, this means we could have expected another 7 survey answers.
We decided to not re-run the survey for those 32 individuals, as it would add another threat to validity, namely opinions having changed compared to the original cross-section.

\subsubsection{Reliability}

Reliability describes the degree to which similar results would be obtained if the same study would be repeated, by the same or by other researchers.

While we described the study designs, data collection, and data analysis procedures in detail, there is a subjective element to several parts of our studies.
Specifically, the hypotheses definition process is based on our experiences, understanding of contemporary SE research methods, and belief systems about ethics in recruiting student subjects.
Similarly, the qualitative analysis at several parts of our studies is subject to the researcher's interpretation.
Specifically, we categorised the study type in the mapping study and applied open coding to the survey free-text answers.
Both were not validated.
For the study type, we tried to be conservative in our categorisation and rely mainly on verbatim statements from the papers.
For instance, we did not try to judge whether a study is indeed a ``controlled experiment'', or would be better categorised as a ``quasi-experiment'' or even a ``field experiment''.
Given the maturity and the lack of common terminology with respect to research methodology in SE, we believe that a certain lack of precision is acceptable here.
For the open coding of survey answers, we take an interpretivist viewpoint that subjectivity and different interpretations among multiple coders are in fact desirable.
Hence, reliability is indeed not given, but also not desirable. \section{Results}
In the following, we will discuss the results of our systematic mapping study and of the survey.
Plots for mapping study data are displayed using grey colours, while light blue is used for survey data.

\subsection{Systematic Mapping Study}
\label{sec:resMap}
The final selection of papers included 372 primary studies published in top SE venues.

\begin{figure}[ht]
	\centering
	\includegraphics[width=120mm]{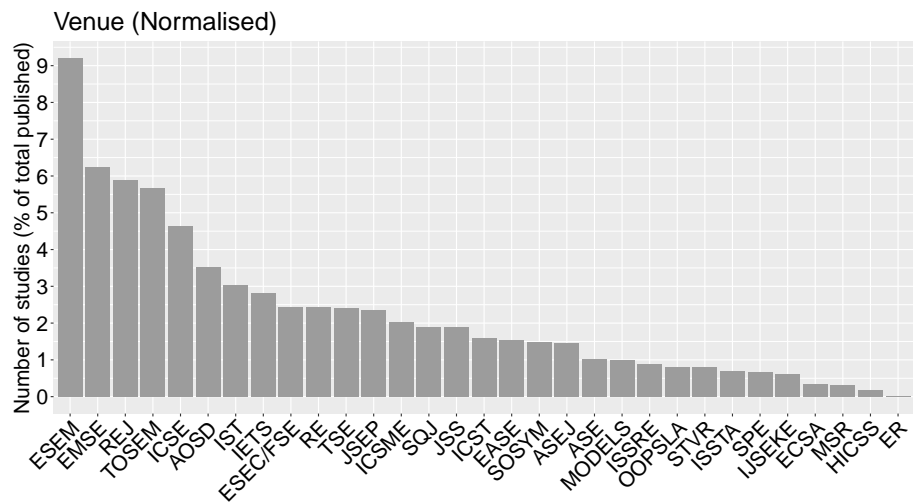}
	\caption{Venue of Mapping Study Papers (Normalised).}
	\label{fig:slrVenuesNorm}
\end{figure}

The paper count per venue is depicted in Figure~\ref{fig:slrVenuesNorm} relative to the total number of papers published at each venue.
On the lower end of the scale, we see venues that focus on automated tasks and therefore have a naturally low percentage of studies with students, e.g., MSR and ASE, and studies with incomplete indexing in the used databases (as discussed in the threats to validity), e.g., HICSS and IJSEKE.
The list is topped by the two venues focusing on empirical SE research, EMSE and ESEM conference.
Similarly, it can be seen that the flagship conferences and journals, i.e., EMSE, TOSEM, ICSE, ESEC/FSE, and TSE, all have a percentage of student studies over the average.
Finally, while a number topic-specific SE venues have high percentage of publications with student studies, e.g., REJ and RE, others have much lower percentage, e.g., MODELS and ISSRE.

\begin{figure}[ht]
	\centering
	\includegraphics[width=100mm]{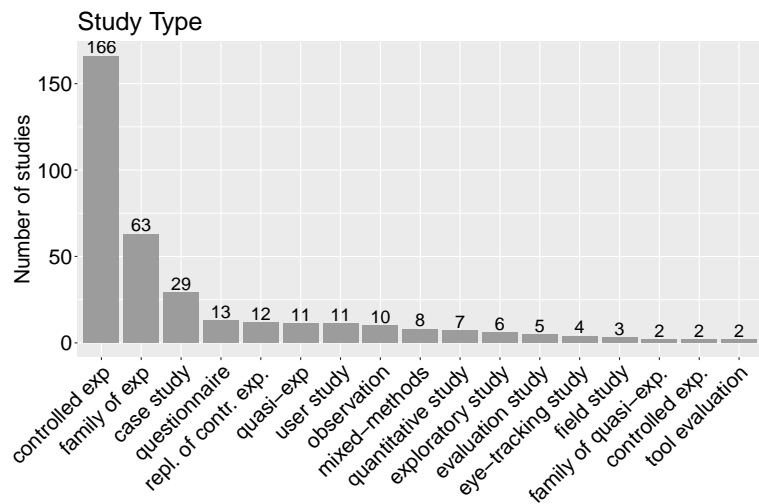}
	\caption{Study Types of Mapping Study Papers.}
	\label{fig:slrStudyTypes}
\end{figure}

In Figure~\ref{fig:slrStudyTypes}, the research strategies applied in the primary studies are depicted (excluding strategies that appeared in only one primary study).
The dominant strategies are controlled experiments (166 studies) and families of controlled experiments (63), followed by case studies (29 publications), surveys/questionnaires (13 publications), replications of controlled experiments (12), and individual quasi-experiments (11).
A number of different strategies exist with 10 or less publications, e.g., observational studies and mixed-method studies.

\begin{figure}
	\centering
	 \subfloat[Sample Size.]{{\includegraphics[width=55mm]{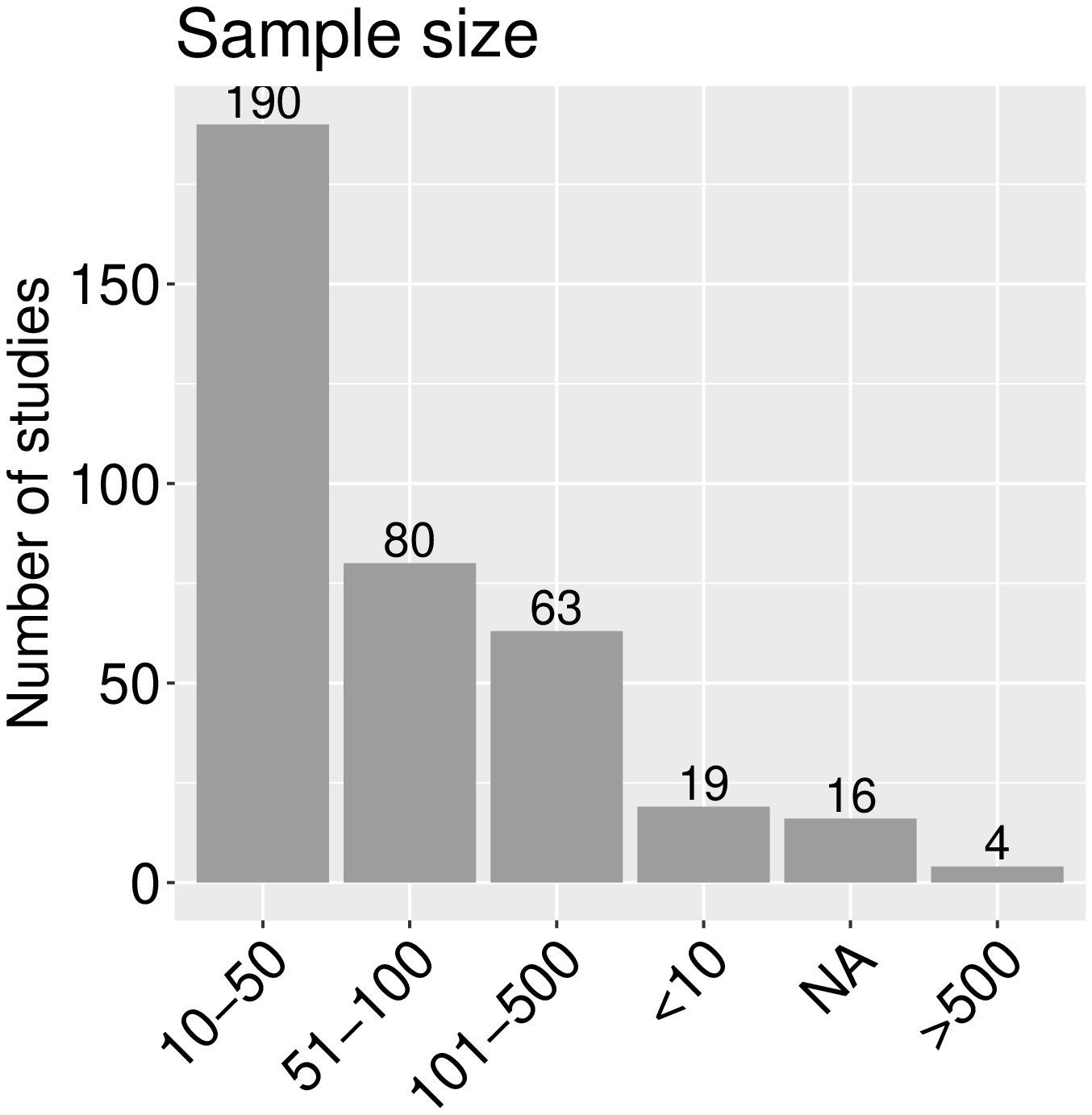}}}
	 \qquad
	\subfloat[Compensation of Students.]{{\includegraphics[width=55mm]{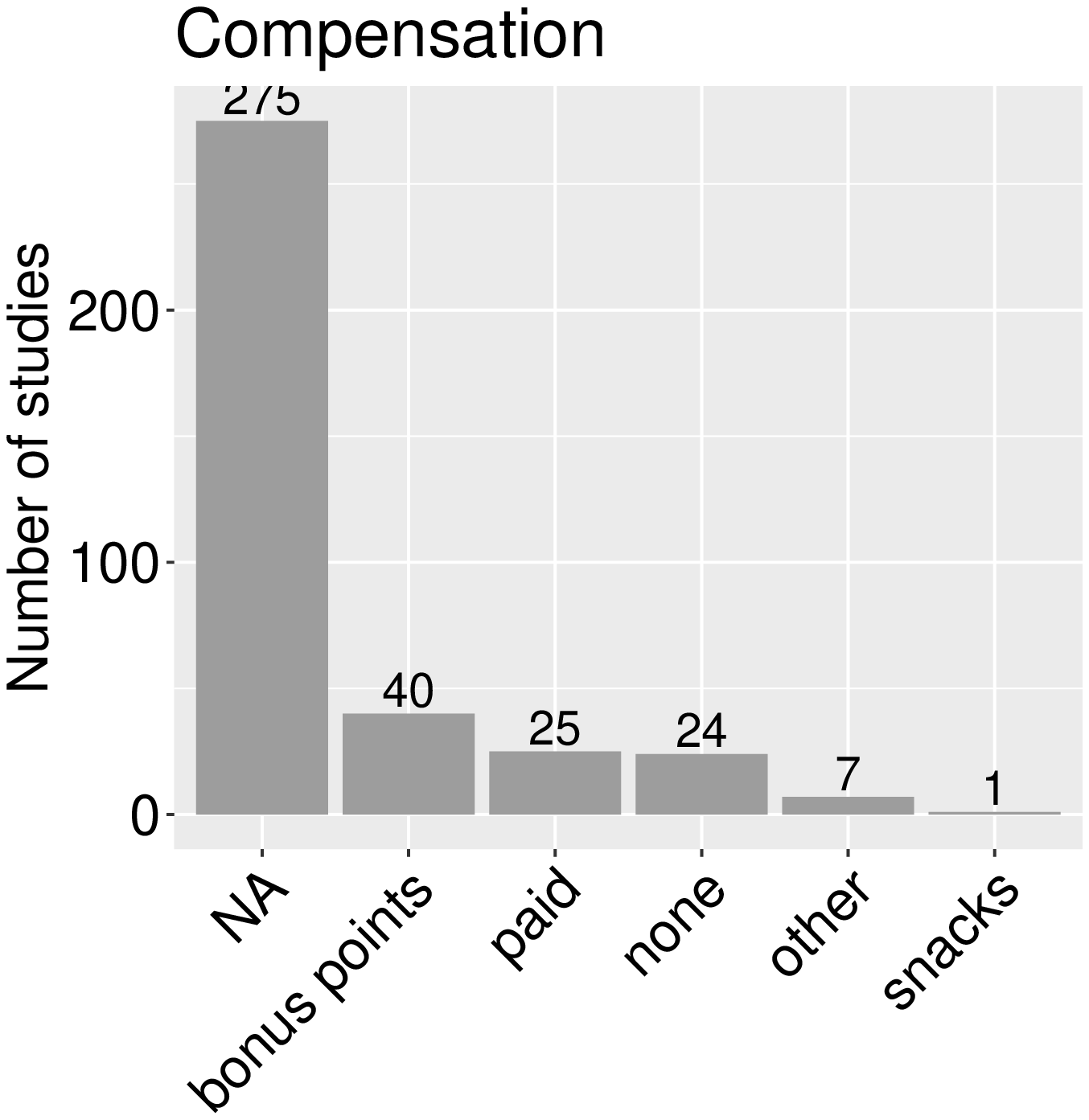}}}
\caption{Sample Size and Compensation in Mapping Study Papers.}
\label{fig:sampleXcompensation}
\end{figure}

On the left-hand side of Figure~\ref{fig:sampleXcompensation}, the sample size of the primary studies is depicted.
Approximately half of the studies (190) have between 10 and 50 subjects.
This is followed by larger studies, i.e., 80 studies with between 51 and 100 students and 63 studies with between 101 and 500 students.
Only 4 publications have a sample size of over 500.
16 publications do not clearly name their subject count.
Finally, 19 publications have  less than 10 subject.

Sample size is directly related to the research strategy.
Depending on the type of data collected and on how the actual study is conducted, sample sizes may vary considerably.
For instance, running and analysing a survey with a large number of subjects is typically considerably less work than conducting an observational study with the same number of subjects.
Furthermore, we only recorded the number of student subjects.
A study might therefore have a sample size that is larger than the one we name here, in case non-student subjects participated as well.

Of all primary studies, only 22 primary studies clearly stated that they obtained ethics approval.
In stark contrast, 347 studies did not mention ethics approval at all.
Finally, 3 studies named that they did not have to obtain ethics approval, but clarified that there are central guidelines without a formal approval process, e.g., mandated by the university or the country/state.

The majority of studies recruited students on a voluntary basis (154 publications).
77 publications did not state how subjects were recruited.
Similarly, 107 did not state whether or not participation was voluntary, but stated that the study was part of a course.
We can therefore not tell whether students participated on a voluntary basis or not.
Finally, 34 studies had mandatory participation by students enrolled in a course.
We included in this category studies that were performed on the basis of graded assignments or exams, since it can be argued that participation is mandatory if the student wants to attain a good or excellent grade.

The right-hand side of Figure~\ref{fig:sampleXcompensation} depicts how students were compensated in the primary studies.
Approximately 75\% of the studies (275) did not state how or if students were compensated for their participation.
Of the remaining 25\%, 40 studies compensated students in the form of bonus points, 25 with a financial reward, 1 study with snacks, and 7 studies or in another form.
Finally, 24 studies explicitly stated that there was no compensation in addition to the practice and training acquired through the study participation.

\begin{figure}[ht]
	\centering
	\includegraphics[width=55mm]{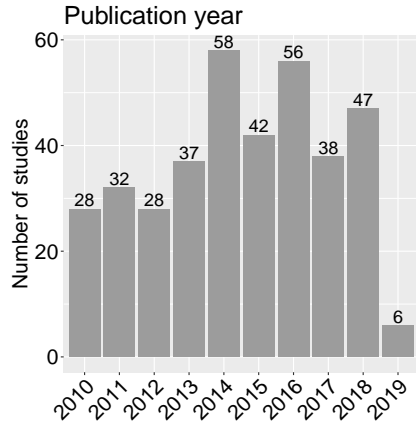}
	\caption{Papers per Publication Year.}
	\label{fig:slrPapersPerYear}
\end{figure}

Looking at trends over time, the overall publication count per year is depicted in Fig.~\ref{fig:slrPapersPerYear}.
Note that 2019 is incomplete, as the data was extracted in early April 2019.
While there are slightly higher numbers, on average, from 2014 on, there is too little data to consider this a trend in the field.
In particular, 2014 and 2016 stand out.

\begin{figure}[ht]
\centering
\includegraphics[width=\columnwidth]{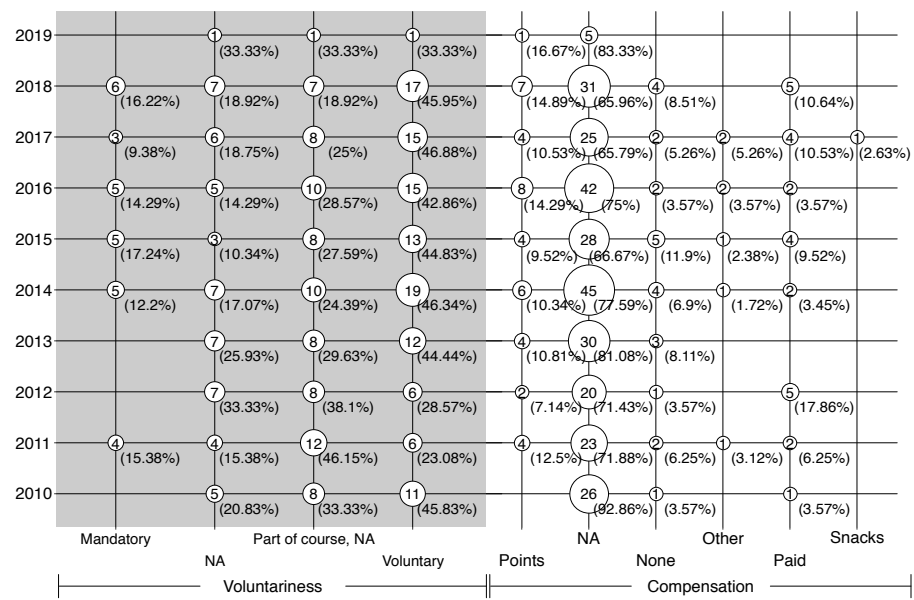}
\caption{Voluntariness and Compensation in Studies Per Year.}
\label{fig:slrBubbleYearVolComp}
\end{figure}

Going in more depth, we plotted the voluntariness and the compensation on a yearly basis in Fig.~\ref{fig:slrBubbleYearVolComp}.
The y-axis depicts the years in increasing order, and the x-axis depicts the voluntariness (grey-shaded, left-hand side) and the compensation (no shade, right-hand side).
Each bubble shows the number of studies for a given year with the reported condition.
Additionally, next to each bubble the percentage of studies reporting the given condition is depicted for each given year.
For example, the bubble at the intersection of 2011 and ``Mandatory'' shows that 4 studies, or 15.38\% of studies in that year, reported mandatory participation.
Similar to the overall numbers, voluntary participation is the most common study option in all years, followed by studies that were part of a course (with no further details on voluntariness).
The same pattern is visible in the compensation plot, with most studies in all years not specifying the compensation.
Looking at the progress over time, there is no clearly visible trend in either of the plots.
For instance, the percentage of studies with voluntary participation is almost identical in 2018 and 2010, while dropping in 2011 and 2012.
Similarly, while the percentage of studies not reporting compensation is highest in 2010, there is no decrease over time.

\begin{figure}[ht]
\centering
\includegraphics[width=\columnwidth]{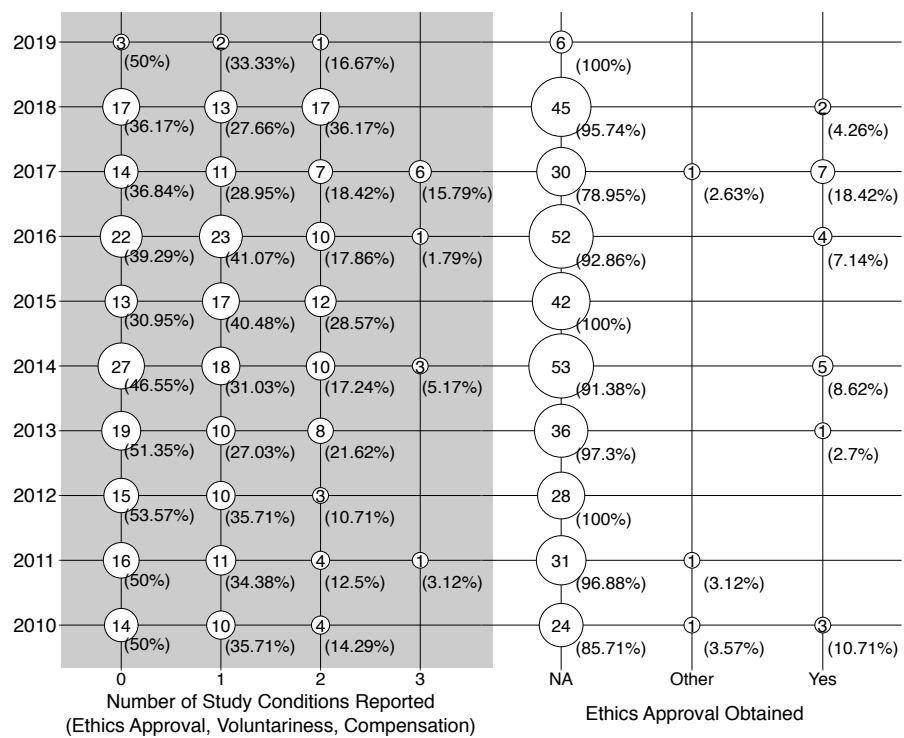}
\caption{Number of Study Conditions and Ethical Approval Status in Studies Per Year}
\label{fig:slrBubbleYearNumEthics}
\end{figure}

In Fig.~\ref{fig:slrBubbleYearNumEthics}, we depict an additional bubble plot plotting reported study conditions against the publication year.
The x-axis depicts how many of the three study conditions (voluntariness, compensation, ethics approval) were reported by the different publications (grey-shaded, left-hand side) and the ethics approval (no shade, right-hand side).
There is a visible decrease over time of studies not reporting any of the three conditions.
While between 46.55\% and 53.57\% lacked these conditions in between 2010 and 2014, the numbers are substantially lower afterwards, ranging between 30.95\% and 30.29\%.
While 2019 has again 50\% of studies not reporting any condition, this data is incomplete, as discussed.
Overall, only 11 papers report all three conditions, 1 in 2011, 3 in 2014, 1 in 2016, and 6 in 2017.
Due to this low number, we decided to reference these papers in full in Table~\ref{tb:completeStudies}.
For ethics approval, the percentage of non-NA answers is so low that any analysis of trends would likely be arbitrary and due to random error.

\begin{table}
\caption{Primary Studies Reporting all of Voluntariness, Compensation, and Status of Ethics Approval.}
\label{tb:completeStudies}
\begin{tabular}{|p{0.05\textwidth}|p{0.25\textwidth}|p{0.08\textwidth}|p{0.15\textwidth}|p{0.17\textwidth}|p{0.1\textwidth}|}
\hline
Key & Study Type & Sample Size & Voluntariness & Compensation & Ethics Approval\\
\hline \hline
\cite{anvari17} & Mixed-methods & 75  & voluntary & Other & Yes \\
\hline
\cite{tu16} & Controlled experiment & 48  & voluntary & Movie vouchers & Yes \\ \hline
\cite{dahan14} & Controlled experiment & 53  & voluntary & Bonus points & Yes \\ \hline
\cite{floyd17} & Controlled experiment & 35  & voluntary & Paid & Yes \\ \hline
\cite{barik17} & Eye-Tracking study & 56  & voluntary & Bonus points & Yes \\ \hline
\cite{sturm14} & Controlled experiment & 60 & voluntary & Bonus points & Yes \\ \hline
\cite{sakhnini17} & Controlled experiment & 45  & voluntary & Paid & Yes \\ \hline
\cite{riaz17} & Family of cont. exp. & 155  & mandatory & Bonus points & Yes \\ \hline
\cite{grubb17} & Controlled experiment & 15  & voluntary & paid & Yes \\ \hline
\cite{budgen11} & Quasi-Experiment & 20  & mandatory & none & Other \\ \hline
\cite{riaz14} &  Controlled experiment & 50  & voluntary & none & Yes \\
\hline 
\end{tabular}
\end{table}

\subsection{Survey}
We received 100 answers to our survey\footnote{Due to this convenient number, we use number of answers and percentage interchangeably in the following.}, corresponding to a response rate of 21.1\%.
One participant did not answer any questions beyond the demographic questions.
Furthermore, two participants left the last block of questions regarding their opinions on study circumstances unanswered.

\subsubsection{Demographics}
Our participants are mainly located in United States of America (13 participants), Germany (11 participants), Spain (10 participants), and Sweden (9 participants).
The remaining countries follow at some distance, with Israel having 5 participants, Australia, Brazil, Canada, Italy, the Netherlands, and Portugal having 4 participants each, Finland, New Zealand, and Switzerland having 3 participants each, and Austria, Poland, and the United Kingdom having 2 participants each.
Finally, 9 further countries\footnote{We do not list countries with only one participant, to ensure participant anonymity.} had one participant each, 3 participants did not state their location, and one answer was not identifiable (a free-text answer that could not be mapped to a country).

74\% of participants have a PhD degree as their highest degree, 12\% a habilitation degree, and 10\% of participants hold a Master degree.
3 participants hold a high school degree and 1 participant another degree.

Regarding currently or last held academic position, 14 participants stated professor, 9 full professor, 24 associate professor, 22 assistant professor, 15 PhD student/candidate, 5 postdoc, and the remaining 10 various other positions (4 senior lecturer, 2 senior researcher, 1 research scientist, 1 master student, and 2 adjunct professor).

\subsubsection{Study Conditions}
Of our participants, 43\% stated that they require ethics approval.
17\% do not require ethics approval, but have to follow mandatory steps that regulate how studies with student participants are to be conducted.
Finally, 39\% are neither required to obtain ethics approval, nor to follow mandatory steps.

\begin{figure}[ht]
	\centering
\subfloat[Number of Studies in Last 5 Years]{{\includegraphics[width=.45\columnwidth]{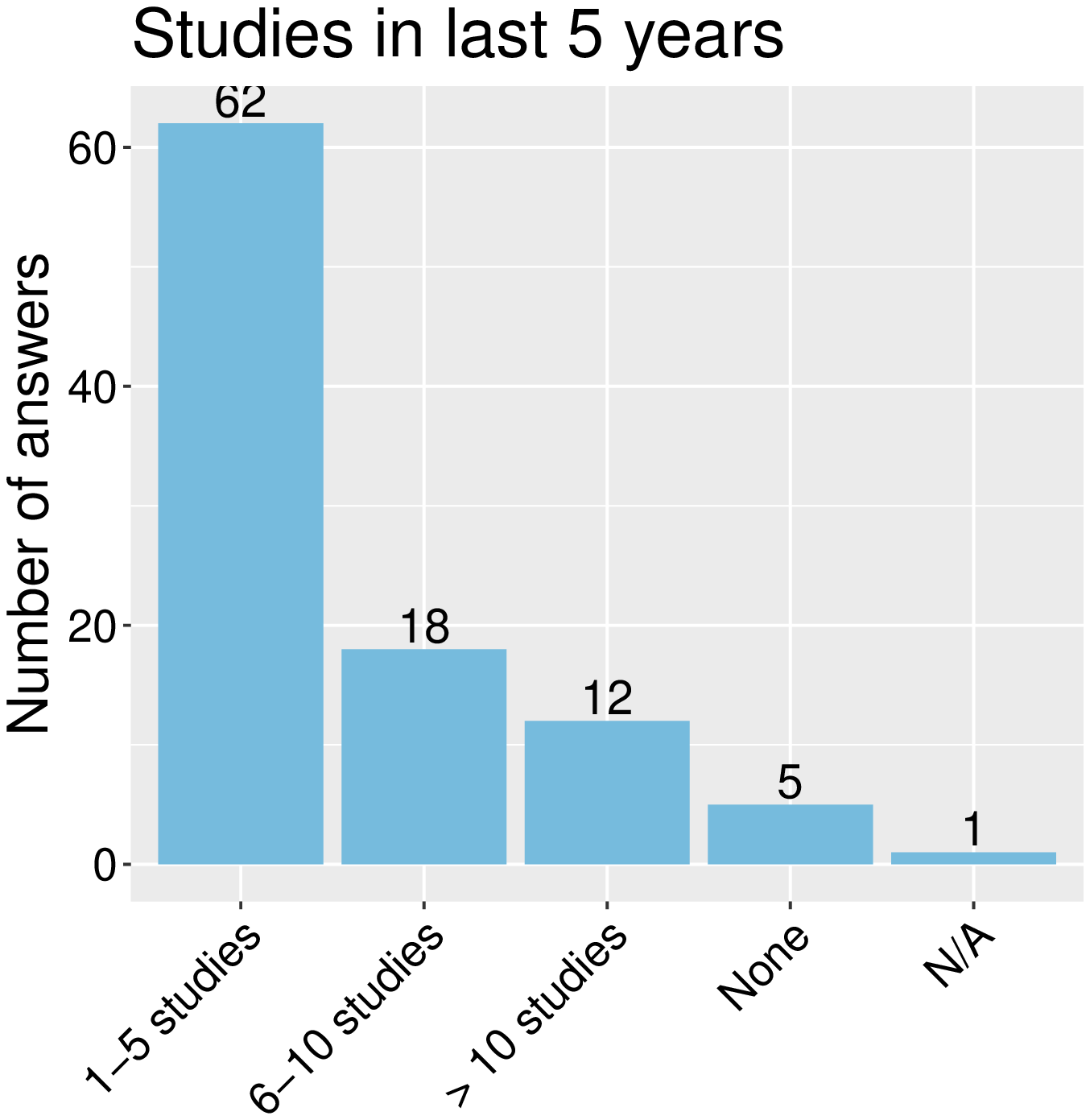}}}
	\qquad
	\subfloat[Recruitment of Own Students]{{\includegraphics[width=.45\columnwidth]{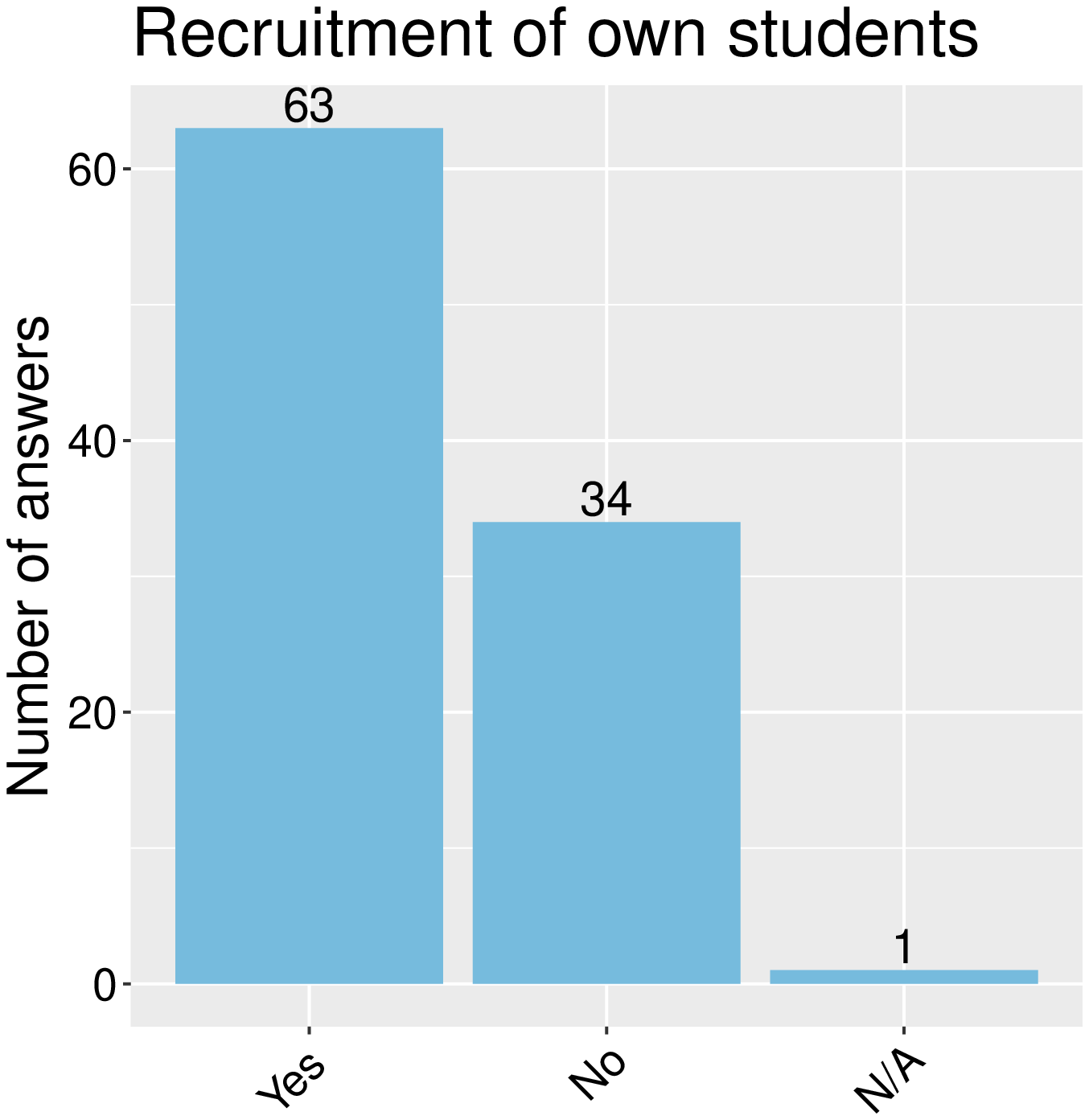}}}
	\caption{Number of studies and Recruitment of own students}
	\label{fig:surNumxRecruit}
\end{figure}

In Figure~\ref{fig:surNumxRecruit} (left-hand side), we depict the number of student studies participants conducted during the last 5 years.
The majority conducted 1 to 5 studies (62\%), followed by 6 to 10 studies (18\%) and more than 10 studies (12\%).
Finally, 5 participants did not conduct any studies with student participation, and one participant did not answer.

During those studies (Figure~\ref{fig:surNumxRecruit} (right-hand side)), 63\% of participants recruited their own students, and 34\% did not do so.

65\% of our participants used entirely voluntary participation.
30\% offered students a choice to participate, but participation influenced grading in some way.
Finally, 2 participants used mandatory participation.

\begin{figure}[ht]
	\centering
	\includegraphics[width=100mm]{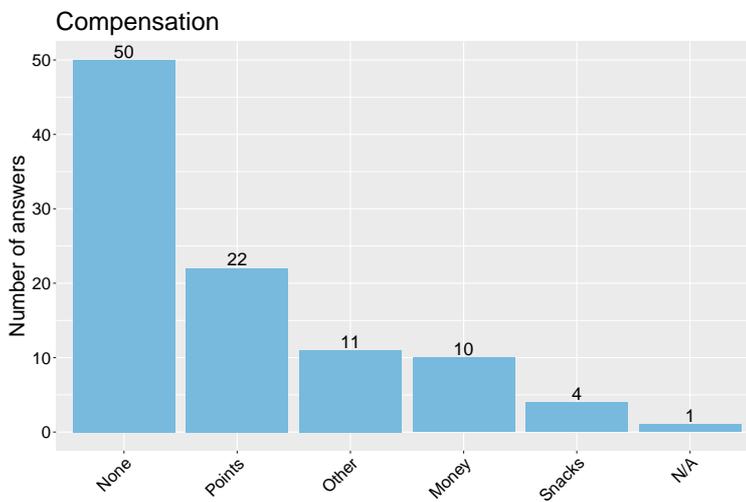}
	\caption{Subject Compensation in the Studies}
	\label{fig:surCompensation}
\end{figure}

In Figure~\ref{fig:surCompensation}, we depict how the participants compensated participation in their studies.
50\% did not offer any compensation to their study participants.
22\% offered bonus points/credits in their course.
10\% offered a financial reward, including lotteries with cash prizes.
Finally, 4\% offered snacks and 11\% other forms of compensation.
Regarding the bonus point compensation, many studies did not provide further details on how exactly these points or credits were offered, while some papers stated that the bonus points would count towards the final grade.
From personal experience, we also know of cases where bonus points lead to more than the ``full points'' in the course.
For instance, a course could have a total of 100 points in all assignments and exams, but the bonus points would make it possible to reach 110.

The vast majority of participants, or 88\%, use informed consent in their studies.
9\% did not use informed consent and 1 participant did not answer.

Similarly, the majority (92\%) offered their study participants to withdraw at any time, while 5\% did not offer this option.

\subsubsection{Opinions on Study Conditions}
In addition to asking participants how they conducted studies in the last years, we also asked them to state their opinions on a number of statements.
These relate to how studies should be connected to a course, what the participation conditions should be, and how acceptable different practices are.
Finally, we asked them which characteristics of a study should be reported in a publication.
In all following figures, the green bars to the right of the centre line depict agreement (light green) and strong agreement (dark green), the grey bars in the middle depict neutrality towards the statement, and the brown bars to the left of the centre line depict disagreement (light brown) and strong disagreement (dark brown).
The statements are ordered by overall agreement.
In addition to the overall response, we also report numbers for the sub-groups of students (PhD candidates/students, master students, and research scientist, $n=17$), early-stage researchers (PostDoc, Assistant professor, senior/university lecturer, and adjuncts, $n=35$), and senior faculty (Associate professor, professor, full professor, $n=48$).

\begin{figure}[ht]
	\centering
	\includegraphics[width=120mm]{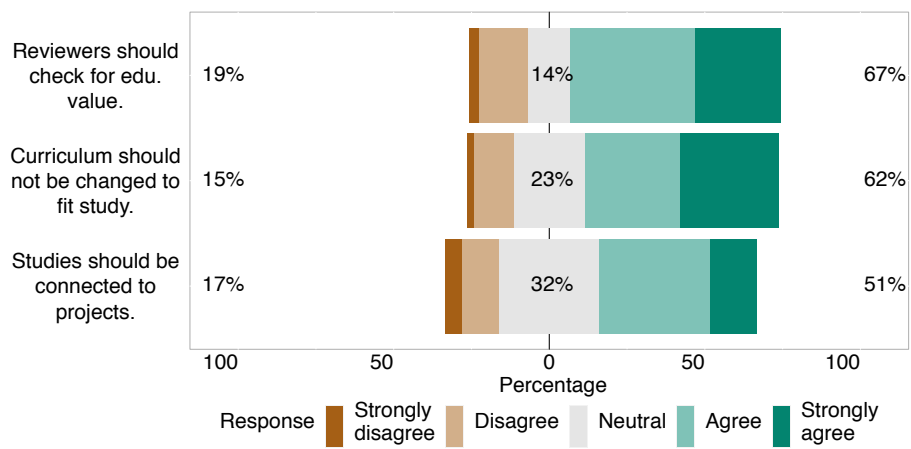}
	\caption{Opinions on the Study-Course Connection.}
	\label{fig:surCourse}
\end{figure}

Figure~\ref{fig:surCourse} depicts the participants' opinions regarding the connections between study and a course.
Overall, there is (strong) agreement that reviewers of a manuscript should specifically check whether or not the study had educational value to the participants, that the curriculum should not be changed to fit a study into the course, and that studies should be connected to course projects.
The latter point is related to projects in particular, i.e., practical course moments.
Disagreeing with this statement does not necessarily mean that the study is completely disconnected from the course content, it might simply not be connected to a practical course moment.

The sub-groups do not exhibit large differences here for this block of questions.
Students have less disagreement on the statement that studies should be connected to projects (7\% compared to 21\% in early-stage researchers and 18\% in senior faculty).
Similarly, the student group has a more neutral opinion (38\% compared to 18\% and 22\%) and less disagreement (6\% compared to 15\% and 17\%) on the statement that the curriculum should not be changed to fit in a study.

\begin{figure}[ht]
	\centering
	\includegraphics[width=120mm]{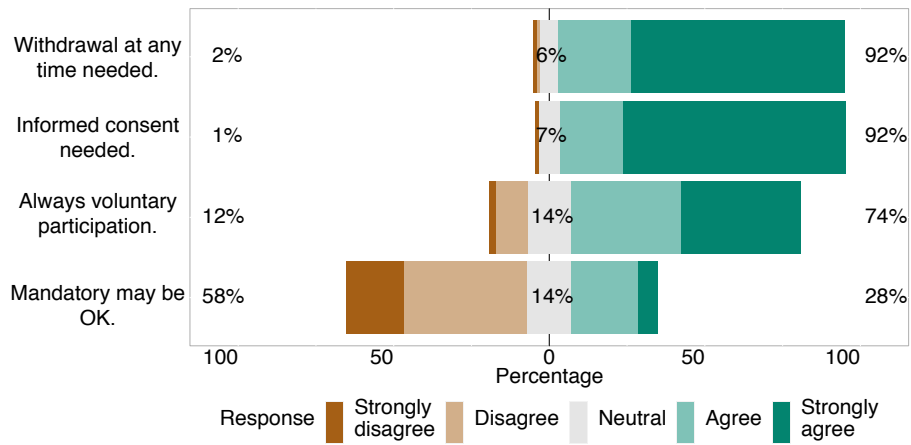}
	\caption{Opinions on Study Circumstances.}
	\label{fig:surCircumstances}
\end{figure}

Figure~\ref{fig:surCircumstances} depicts the participants' opinions regarding the circumstances of the study.
There is strong agreement that informed consent including the possibility to withdraw at any time should be used.
Similarly, the majority of participants agree that studies should always be voluntary, and disagree that mandatory attendance may be warranted.

Interestingly, 27 participants agreed or strongly agreed that mandatory participation may be warranted.
Among those, 16 are senior faculty (associate professor, professor, full professor), 5 are early-stage researchers (assistant professor, senior lecturer), and the remaining 6 students (PhD student/candidate, research scientist).
In terms of geographic distribution, 14 European participants believed mandatory participation might be warranted, followed by 5 from North America, 4 from Oceania, and 3 from Asia.
Several participants clarified their answers in free text.
For instance, one participant noted that, in a course that is elective, mandatory participation might be warranted within limits.
That is, the students should still have an alternative, e.g., completing x out of y assignments, where one option is participation in the empirical study.
Six people noted that participation is different from data usage.
That is, the instructor may require participation in a study, e.g., to expose the students to the process of empirical studies, but opting out of their data being used for research purposes should always be possible.

In terms of sub-groups, there is very little difference.
The only notable exception are the views on mandatory participation.
Here, only 14\% of early-stage researchers agree, with 71\% disagreeing.
In comparison, senior faculty have 35\% agreement and 48\% disagreement, and students have 35\% and 59\%.

\begin{figure}[ht]
	\centering
	\includegraphics[width=120mm]{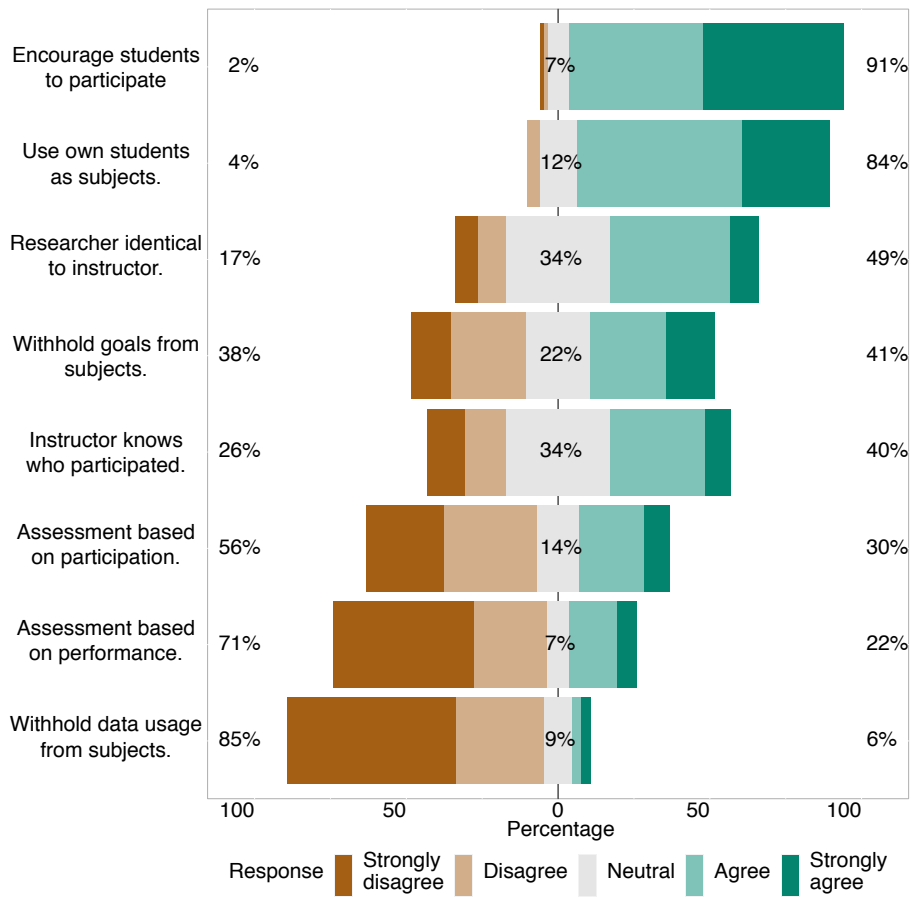}
	\caption{Opinions on Acceptability of Practices}
	\label{fig:surAcceptability}
\end{figure}

Figure~\ref{fig:surAcceptability} depicts the participants' opinions regarding a number of practices that are discussed in related work on ethics in student studies.
The highest agreement (91\%) was registered on the practice to encourage students to participate in a study.
Only 2\% disagreed with this practice.
84\% agreed that it is acceptable to use their own students (registered in a course given by or supervised by the researcher) as subjects in an empirical study.
Only 4\% disagreed, while 12\% remained neutral.
49\% agree that it is acceptable for the researcher conducting the study to be the same person as the course instructor, while 17\% disagree and 34\% remain neutral.
The next two statements caused a considerable polarisation in answers:
41\% agree and 38\% disagreed that it is acceptable to withhold study goals prior to a study.
Part of this polarisation is explained by free-text answers of participants stating that it is sometimes necessary to withhold the detailed hypotheses in order to not risk the validity of the studies.
Similarly, 40\% agree and 26\% disagree that it is acceptable for the course instructor to know who participated in a study, while 34\% stated a neutral opinion.
The remaining three statements all have considerable disagreement from the survey participants:
56\% disagree that it is acceptable to base part of the assessment in a course on participation in a study, with 30\% agreeing and 14\% neutral.
71\% disagree, most of which strongly, that it is acceptable to base part of the assessment in a course on performance in a study, with 22\% agreeing and 7\% neutral.
Finally, 85\% disagree that it is acceptable to withhold information regarding the use of data collected in a study from the students.
Only 6\% agree with this statement and 9\% remain neutral.

As before, we do not see large disagreements between the sub-groups.
One difference is that students disagree substantially more on the statements that it is acceptable to withhold goals (12\% agreement, 53\% disagreement) compared to senior faculty (44\% agreement, 29\% disagreement) and early-stage researchers (52\% agreement, 42\% disagreement).

\begin{figure}[ht]
\centering
\includegraphics[width=\columnwidth]{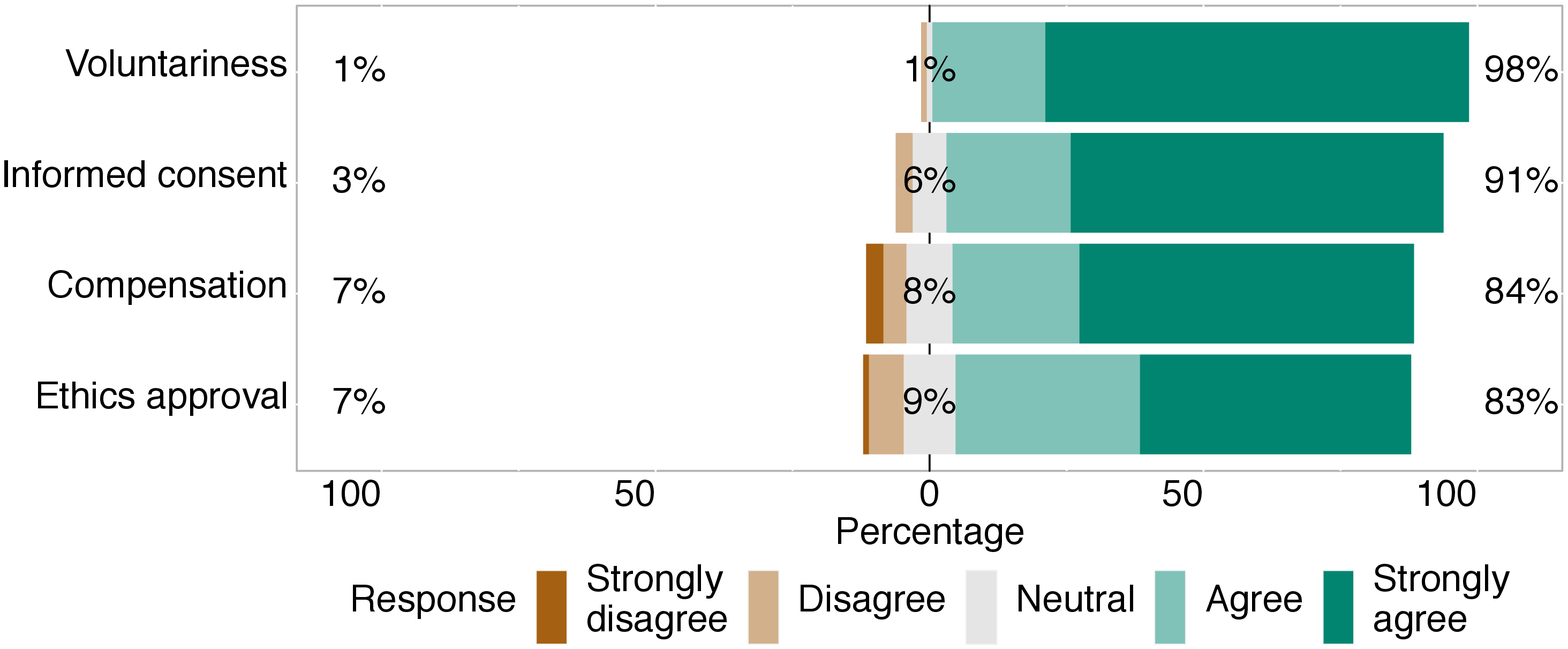}
\caption{Opinions on Reporting of Study Information}
\label{fig:surReporting}
\end{figure}

The final set of statements relate to reporting of study characteristics in a publication manuscript.
All statements have very high agreement, as depicted in Figure~\ref{fig:surReporting}, generally favouring that study characteristics should be stated transparently.
98\% agree that voluntariness of participation should be reported, while the use of informed consent (91\% agreement), student compensation (84\% agreement) and status of ethics approval (83\% agreement) received slightly less agreement.

As the overall picture is very decided, the sub-groups do not exhibit any large differences in any of the points.

\subsubsection{Hypotheses Evaluation}
Overall, these results allow us to evaluate our hypotheses formulated prior to the survey.
An overview of the evaluation is given in Table \ref{tb:hypResult}.

\begin{table}
\caption{Hypotheses Evaluation}
\label{tb:hypResult}
\begin{tabular}{|p{0.05\textwidth}|p{0.4\textwidth}|p{0.08\textwidth}|p{0.35\textwidth}|}
\hline
ID & Hypothesis & Yes/No & Key Metric\\
\hline
$H_1$ & The majority of participants are not required to obtain ethics approval
        & No & 43\% require ethics approval\\
\hline
$H_2$ & The majority of participants recruit their own students for empirical studies. & Yes & 63\% recruit own students\\
\hline
$H_3$ & The majority of participants perform voluntary studies, as a part of courses. & Yes & 95\% have voluntary participation \\
\hline
$H_4$ & The majority of participants do not offer compensation to their subjects. & ? & 50\% do not offer compensation, 50\% offer some form of compensation\\
\hline
$H_5$ & The majority of participants use informed consent, including the option to withdraw voluntarily. & Yes & 88\% use informed consent, 92\% the option to withdraw.\\
\hline
$H_6$ & The majority of participants agree that studies should relate to course learning outcomes or project work. & Yes & Over 50\% agreement on all related points.\\
\hline
$H_7$ & The majority of participants agree that informed consent and withdrawal should be offered. & Yes & 92\% agree \\
\hline
$H_8$ & The majority of participants disagree that information on ethics approval, voluntariness, compensation and informed consent should be included in publications. & No & Between 83\% and 98\% agreement \\
\hline
\end{tabular}
\end{table}

$H_{1}$, that the majority of participants are not required to obtain ethics approval, is not supported by the survey data.
43\% of all participants require ethics approval, and 17\% have at least mandatory steps to follow.
Looking at individual countries, we get a mixed picture.
For several countries, all participants answered that ethics approval is required, indicating national legislation.
Among these are, e.g., the Netherlands, New Zealand, Canada, and Israel.
In other countries, e.g., Brazil and Spain, the answers were mixed with no option clearly dominating.
In the country with the largest participation, the USA, 11 participants stated that they needed ethics approval, one stated that they do not need ethics approval, but have mandatory regulations to follow, and one participant stated that they did neither require ethics approval, nor have to follow mandatory guidelines.
Finally, for some countries, the majority answered that no approval is needed, e.g., nine out of eleven answers by German participants.

$H_2$, that the majority of participants recruit their own students for empirical studies, is supported by our data.
63\% of participants answered that they did so in the past five years.
Similarly, 84\% agree or strongly agree later in the survey that it is acceptable to do so.

$H_3$, that the majority of participants perform voluntary studies, as a part of courses, is also corroborated by our data.
Overall, 95\% of participants answered that their studies were voluntary.
Of the 65 participants that recruited their own students, 61 answered that the recruitment was voluntary.

For $H_4$, the majority of participants do not offer compensation to their subjects in addition to the learning benefit, the answer is mixed.
While no compensation is the most common answer (with 50 participants), there are a variety of other compensations being offered, also summing up to 50\% of the participants.

$H_5$, that the majority of participants use informed consent, including the option to withdraw voluntarily, receives by far the strongest support from our survey data.
88\% of participants used informed consent, and 92\% offered students the option to withdraw from the study.

$H_6$ targets the different opinions participants have regarding the relation of a study to the course curriculum and learning outcomes.
All statements have an agreement of over 50\%, thus supporting the hypothesis.
However, it is interesting to note that they also have reasonably high disagreement values.
For instance, 19 participants disagree that reviewers of an empirical study involving students should check for educational value of a study.
No free-text answers were left explaining these points.

For $H_7$, we expected that the majority would agree that informed consent and option to withdraw need to be present in a study.
This was clearly supported, with 92\% agreement to both statements.
The second part of the hypothesis aimed towards mandatory participation, stating that the majority would find mandatory participation acceptable given the circumstances.
However, 58\% disagreed that mandatory participation may be OK.

Finally, hypothesis $H_8$ considered different characteristics of a study to be included in publications.
Given the lack of information on ethics approval, voluntariness, compensation, and informed consent in our mapping study, we expected most participants to answer that this information is not necessary.
On the contrary, we found large support that these aspects should in fact be described in a study, with 98\% agreeing that voluntariness should be reported, followed by informed consent (91\%), compensation (84\%), and ethics approval (83\%).
This is clearly in contrast to the findings of our mapping study. \section{Discussion}
Several of our findings lead to interesting discussions regarding the ethics of empirical studies with student participation in SE.
We will discuss these in the following according to the four pillars of ethical research by \cite{singer02,vinson08}: scientific value, confidentiality, informed consent, and beneficence.

\subsection{Scientific Value}
Primary studies with student participation are published substantially in the top SE venues, clearly indicating that there is \emph{scientific value} in many of these studies.
In our mapping study, there are not publication trends that clearly stand out.
However, it is worth noting that the prime venues such as ICSE or ESEC/FSE do not exhibit any patterns that suggest an inherent bias against studies with student subjects.
Instead, these venues have among the highest percentage of such studies.

The primary method used in empirical studies with students is controlled experiment, along with related methods, e.g., families of controlled experiments, (families of) quasi-experiments, and replications of controlled experiments.
This is sensible given that student populations are often homogeneous in terms of knowledge and experience, there is control as to which techniques they have learned, and recruitment is easy compared to professional subjects.
Furthermore, in qualitative methods such as case studies, observational studies, and ethnography, the case context is a deciding factor and can typically not be separated from the studied phenomenon \cite{stol18}.
Therefore, such studies are only feasible with student subjects if their context is of direct interest to the study, e.g., studies on curriculum design, or on student behaviour.
Similarly, it is not unexpected that qualitative studies in general are under-represented, given that the number of qualitative studies is generally lower in SE \cite{storey19}.
Potential factors playing a role here are biases against students as subjects, against qualitative studies in general, and a comparable lack of maturity in qualitative methods compared to quantitative methods in SE.

While ideally sample size should be appropriate according to the chosen research strategy and context, reviewers might tend to reject papers with lower sample sizes, even if justified.
However, we see no such indication in our data.

\subsection{Confidentiality}
Regarding the issue of \emph{confidentiality}, we do not either see a need for action.
The majority of survey participants find it unacceptable to withhold information regarding data usage from student subjects, and find it acceptable to withhold goals and hypotheses only if needed for validity reasons.
Similarly, recent years have seen increasing awareness regarding privacy laws such as the EU General Data Privacy Regulation (GDPR) \cite{voigt17}, leading in turn to an increased attention to confidentiality issues.
This is also witnessed in our survey by a number of free-text comments stating that it would be illegal to withhold information on how data is used from the student subjects.

\subsection{Informed Consent}
\emph{Informed consent} is used by 92\% of our survey participants, including the right to withdraw from a study.
We did not extract this data from the mapping study, but consent was one of the keywords we searched for when extracting data.
Our feeling is that, while more prominent than ethics approval, use of informed consent was not typically reported in primary studies.
That is, while it is an encouraging sign that the majority of survey participants use informed consent, we see a need for reporting this in the final papers.
As discussed in Singer and Vinson~\cite{singer02}, the elements that should be part of informed consent are disputed in ethics research.
We would like to reiterate the authors' statements that important parts in informed consent are information on voluntariness, the right to withdraw from the study at any time, and the act of consenting to the study.
Guidance can be offered by official regulations that list required elements of informed consent, such as \cite{ecfr45}, or by the guide for ethics approvals given in \cite{vinson08}.

Related to giving the consent, the aspect of whether or not participation in a study is actually voluntary matters.
The majority of survey participants used voluntary participation in their studies.
Similarly, the majority of primary studies in our mapping study used voluntary participation, at least if we exclude those that did not report on the voluntariness.
However, the majority of survey participants recruited their own students, and found it acceptable that the researcher is identical to the course instructor.
Finally, 40\% found it acceptable that the instructor knows who participated and 34\% remained neutral on this statement.
Galster et al.~\cite{galster12} even found it helpful that the instructor is the researcher, since students have more trust than in a researcher unknown to them.
There seems to be little concern that students might feel pressured to participate if the researcher is identical to the researcher \cite{singer02,sieber01,storey01}, even if participation is voluntary.
Suggested measures, such as letting a graduate student handle recruitment instead of the instructor, would not add substantial overhead to a study and should therefore be easy to implement.
This is a clear lack of ethical practice in the field, and we therefore see a strong need for increasing discussion (and corresponding action) regarding our recruitment practices and study conditions.
In addition to the ethical aspect, pressuring students into participating might affect the validity of the findings as well, since students will behave differently compared to voluntary participation.
We would like to believe that only very few researchers knowingly coerce their students into participation, but this pressure can also arise unintentionally.

\subsection{Beneficence}
We believe that achieving a positive \emph{beneficence}, i.e., ``a weighted combination of risks, harms, and benefits to the subjects and society''~\cite{mcneill93}, is the most difficult of the four pillars to address, and the one where the community is lacking most.

As Carver et al.~\cite{carver03,carver10} state, ``students have the right to reach their educational goals, and empirical studies can compete with other instruction forms for the scarce resources in courses, e.g., lecture time''.
On the one hand, students can have a direct benefit of participating in empirical studies, e.g., they can be trained in skills valued in industry, receive training in empirical SE, or receive feedback on their current abilities.
For instance, there is a substantial effort in SE education to expose students to SE trends \cite{cico20}.
Hence, students can familiarise themselves with SE trends in practical course projects, that can be connected to empirical studies.
This will increase their chances on the job market.
On the other hand, there are issues that potentially lower the beneficence, e.g., spending considerable time on an empirical study instead of covering other important course topics, spending time on irrelevant or outdated techniques or methods (as a part of a control group), and higher stress levels due direct or indirect coercion to participate and perform well in an empirical study.
Connecting to the previous example, instructors might decide to investigate whether a new practice or technique increases productivity or quality.
To do so, they necessarily have to compare this new treatment to a  baseline, which means students might end up in the control group being exposed to an outdated method, or at least having to spend considerable time on this method in case of a paired experiment design.
The central issue here is whether or not an instructor is able to objectively assess whether or not their study is justified, and contributes to the students' learning goals.
In particular, confirmation bias might lead instructors to believe that their study contributes more towards the students' learning than it actually does, or that a technique/method developed by themselves is more important or relevant than it actually is. 
In this direction, we also interpret the survey participants' strong agreement on the statement that a curriculum should not be changed to fit in a study.
If the researcher is identical to the instructor, there is a risk that the instructor/researcher mis-judges the relevance of a study topic, especially if studying their own techniques or methods.
Storey et al.'s \cite{storey01} report is a good example of issues that might arise as a result, e.g., exposing students to incomplete prototype tools, or providing students with more support in one area (the area of study) than in another (an alternative assignment).
While this is clearly a difficult topic to address, we believe more discussion is needed surrounding the topic of beneficence to student subjects.

\subsection{Cross-Cutting Concerns}
Finally, as a topic touching on all four pillars, \emph{ethics approval} is an important issue for discussion.
In the mapping study, we see similar statistics as reported for the CHI conference series in \cite{buse11}.
That is, only a small fraction of all studies reports whether they obtained, or if they had to obtain, ethics approval.
In contrast, 83\% of our survey participants believe that ethics approval should be reported.
Hence, there is currently a strong discrepancy between what the community believes should be reported, and what is actually being reported.

We find that only 39\% of our survey participants are required to obtain ethics approval.
This is in conflict to the statement made in Ko et al.~\cite{ko15} that it is likely that authors need ethics approval.
In fact, one participant even pointed out that they would not be able to request approval, since the university does not have processes in place that support this.
From personal experience, we know that reviewers sometimes request ethics approval, or judge it as a negative point if ethics approval was not obtained in a study.
The mixed picture on a per-country basis also clearly shows that legislation in many cases differs between states or universities in the same country, or as some participants pointed out, depends on the agency funding the research.
Therefore, our findings clearly show that reviewers should abstain from requesting ethics approval, unless they know that the authors are indeed required to obtain approval.
However, this does not free the authors from the responsibility to know their local regulations and clarify those - and reviewers might very well request clarification of the regulations surrounding ethics approval.

As to the usefulness of a central board that grants ethics approvals, we see two conflicting opinions.
On the one hand, several participants stated that only the researchers are familiar enough with the method, study context, and the field of study to be able to judge ethical consequences.
Similarly, the experiences by Storey et al.~\cite{storey01} demonstrate that even with ethics approval, ethical issues remain.
On the other hand, one participant pointed out that the value of an IRB is in having an independent unit that is not subject to the same biases that researchers might have towards their own study.
Indeed, we believe that cognitive bias is a central issue in case a researcher has to judge ethical issues surrounding their studies.
In particular, the body of work on confirmation bias impressively demonstrates that human beings are strongly affected in judging and/or justifying their own actions \cite{jorgensen15}.

Without a need for ethics approval, there is a chance that lack of awareness or even disregard for ethical issues can spread, similar to the worrisome picture drawn in 2001 by Hall and Flynn~\cite{hall01}.
Our mapping study results impressively demonstrate that, currently, reporting ethics approval or related regulations is extremely uncommon, and other study conditions are similarly under-reported.
Specifically, the number of studies reporting none of the extracted study conditions (voluntariness, compensation, ethics approval) is alarming.
The bubble plots in Section~\ref{sec:resMap} show no discernible improvement over time, indicating that there is no progress.
This is in direct contrast with the opinions of the survey participants, where a clear majority indicated that study conditions should be reported.
If we consider the survey sample to be representative for the population, we have to conclude that the opinions are not currently reflected in the community's practices.
Therefore, we believe that there is need for action, e.g., by establishing a code of practice, as suggested by \cite{davison01}.
Existing guidelines such as \cite{carver10,vinson08}, and the code of ethics for SE professionals \cite{gotterbarn97} could serve as a basis for such a code of practice for SE research.
We outline the elements of such a code of practice in the following.

\subsection{A Code of Practice for Ethical Studies with Student Subjects}
We believe that a large part of the current shortcomings in SE are due to a lack of reporting practice, something that could be fixed easily.
Specifically, we believe that many important aspects are under-reported because of a lack of attention\footnote{Ironically, we forgot to add information on ethics approval and voluntariness in our survey in the first version of this paper.}
For instance, conference chairs and journals could specifically ask for meta information on study conditions.
Similar disclosures are already common practice in some SE journals for other important aspects, such as financial disclosure or conflict of interest.
For instance, Springer Empirical Software Engineering has a declaration section, where authors are requested to disclose information on funding, conflicts of interest, as well as availability of code and data\footnote{\url{https://www.springer.com/journal/10664/ethics-and-disclosures}}.
The survey also shows that SE researchers familiar with student-subject studies strongly support completely reporting study conditions.

Community agreement is needed on what should be reported, and in what form.
Here, a \emph{code of practice} could help that outlines what elements need to be reported in a study.
As a starting point, we suggest the following elements.
\begin{enumerate}
\item Voluntariness
\begin{itemize}
\item How were the subjects recruited?
\item Was participation in the study mandatory or voluntary?
\item If participation was mandatory, was it possible to opt out of the data being used for research purposes?
\item Were the researchers involved in teaching the course? If yes, how was it ensured that subjects did not feel pressured to participate?
\end{itemize}
\item Compensation
\begin{itemize}
\item How were subjects compensated?
\item If any compensation was given in relation to a university course, e.g., bonus points, which potential effect did non-participation have on the grade?
\end{itemize}
\item Ethics approval and informed consent
\begin{itemize}
\item Was ethics approval obtained? If not, why so (e.g., not possible or necessary)?
\item How was informed consent ensured?
\item Was withdrawal possible at any time? What effect did withdrawal have on compensation?
\end{itemize}
\end{enumerate}

Apart from reporting, there are a number of issues that we believe need to be addressed, but require ongoing discussions and more maturity in the field.
The widely differing opinions in our survey on factors such as voluntariness or compensation, which also showed up to some extent in the review process of this paper, demonstrate clearly that the field has not agreed on such a standard practice.
For instance, while bonus points as a form of compensation might be considered acceptable by faculty and students in some places, students could feel pressured to participate if bonus points are offered.
Similarly, as pointed out by several survey participants, mandatory participation in itself does not need to be unethical, e.g., when it is required to reach certain course learning outcomes.
However, in such a case, students should have the option to decline the use of their data in a research study.
Finally, the experiences of Storey et al.~\cite{storey01} in addition show that obtaining ethics approval alone does not ensure an ethical study, but issues might remain.
Therefore, continued discussions are needed in order to reach a set of accepted practices. \section{Conclusion}
While there is on-going debate on the ethical impact of software products, and the ethics of new research trends such as increasingly capable learning algorithms, there has been a gap in discussion regarding the daily research practices in SE.
Many SE researchers regularly conduct studies with students, as they are easy to recruit, and allow for relatively large sample sizes.
However, students are also vulnerable since they are typically under the control of their instructors, and unethical practices can therefore limit their learning and potentially cause other harm.
Therefore, we aimed to re-visit the current practices in student-subject studies in SE from an ethics perspective.

To do so, we conducted a systematic mapping study on ethical aspects of student use in empirical SE studies.
Based on 372 primary studies in top-SE venues, we find that it is uncommon to disclose information on ethics approval, voluntariness of participation, and compensation.
Out of the 372 primary studies, 160 did not report any, while 125 reported one of the conditions, 76 reported two, and only 11 studies reported all three.

We followed up the systematic map with a survey conducted among the authors of the primary studies, to which we received 100 answers.
The findings show that the majority of participants are not required to obtain ethics approval, that they typically recruit their own students on a voluntary basis, and that student subjects are compensated by approximately half the participants.
Agreement on the acceptability of different practices differs.
The highest approval receive statements that informed consent and the option to withdraw at any time should be given.
Mandatory participation is met by disapproval by the large majority of participants, but 27 participants do agree that it might be warranted under certain conditions, e.g., when students can withdraw the approval for the data to be used for research purposes.
Whether or not the instructor should know student participation receives a mixed response, even though related work on ethics stresses the point that students might feel pressure to participate if this is the case.
Finally, the majority strongly agrees that details on voluntariness, informed consent, compensation, and ethics approval should be disclosed in publications and reviewers should check that this information is present and complete.

Compared to the early 2000s, we see an increased awareness of ethical issues, e.g., reflected in the widespread use of informed consent.
However, the actual figures on reporting these study conditions are alarming and require attention from the community.
Therefore, we believe that a code of practice is required that states which conditions need to be reported when publishing empirical studies with student participation.
Specifically, details on voluntariness of participation, compensation, ethics approval, and the use of informed consent and withdrawal possibilities need to be reported, e.g., as a part of a mandatory appendix in journals, similar to existing disclosure of funding and conflict of interest.
Journal editors and PC chairs then need to make sure that this code of practice is indeed enforced.
In addition, we see the need for discussions on what is ethically acceptable when it comes to studies with student involvement.
From the results of the mapping study, the survey, as well as from the review process of this paper, we can clearly see that opinions on what is ethical practice currently differs substantially.
Finally, beyond committees, regulations, and organisations, we believe that individual researchers in SE need to pay more attention to the ethical aspects of their studies.
Or, as Vinson and Singer put it, ``Ethical research does not happen by chance. Individual researchers must be committed to making their research ethical.''~\cite{vinson08} 
\begin{acknowledgements}
We would like to thank all survey participants, as well as the individuals who gave further input on the survey and feedback on the draft manuscript.
\end{acknowledgements}

\newpage
\bibliographystyle{spmpsci}

\newpage
\appendix

\section{List of Included Venues}
\label{app:venues}
The list of publication venues is as follows:

\begin{table}[!ht]
\caption{Publication Venues with Acronyms}
\label{tab:venues}
\begin{tabular}{|p{0.15\textwidth}|p{0.85\textwidth}|}
\hline
Acronym & Venue \\
\hline
RE & IEEE International Requirements Engineering Conference \\ \hline
SSR & ACM Symposium on Software Reuse \\ \hline
ER & International Conference on Conceptual Modelling \\ \hline
HICSS & Hawaii International Conference on System Sciences \\ \hline
AOSD & Aspect-Oriented Software Development \\ \hline
ICST & International Conference on Software Testing, Verification and Validation \\ \hline
MODELS & International Conference on Model Driven Engineering Languages and Systems \\ \hline
MSR & IEEE International Working Conference on Mining Software Repositories \\ \hline
ISSTA & International Symposium on Software Testing and Analysis \\ \hline
ICSME & IEEE International Conference on Software Maintenance and Evolution \\ \hline
ECSA & European Conference on Software Architecture \\ \hline
ASE & Automated Software Engineering Conference \\ \hline
ISSRE & International Symposium on Software Reliability Engineering \\ \hline
ESEM & International Symposium on Empirical Software Engineering and Measurement \\ \hline
EASE & International Conference on Evaluation and Assessment in Software Engineering \\ \hline
ESEC/FSE & European Software Engineering Conference and the ACM SIGSOFT Symposium on the Foundations of Software Engineering \\ \hline
ICSE & International Conference on Software Engineering \\ \hline
OOPSLA & ACM Conference on Object Oriented Programming Systems Languages and Applications \\ \hline
TSE & IEEE Transactions on Software Engineering \\ \hline
EMSE & Springer Empirical Software Engineering \\ \hline
TOSEM & ACM Transactions on Software Engineering and Methodology \\ \hline
ASEJ & Springer Automated Software Engineering \\ \hline
IST & Elsevier Information \& Software Technology \\ \hline
REJ & Springer Requirements Engineering \\ \hline
SOSYM & Springer Software \& Systems Modeling \\ \hline
SQJ & Springer Software Quality Journal \\ \hline
JSS & Elsevier Journal of Systems and Software \\ \hline
JSEP & Wiley Journal of Software: Evolution and Process \\ \hline
STVR & Wiley Software Testing, Verification \& Reliability \\ \hline
SPE & Wiley Software: Practice \& Experience \\ \hline
IETS & IET Software \\ \hline
IJSEKE & World Scientific Int. Journal of Software Engineering and Knowledge Engineering \\ \hline
\hline
\end{tabular}
\end{table}

\section{Mapping Study: Data Extraction}
\label{app:dataExtraction}
To extract the study conditions and key metrics from the primary studies, we followed the following process.

\begin{enumerate}
\item Check whether paper should actually be included. By default, papers are included in this step. Only exclude if one of the following points applies. Whenever a paper is excluded, note down the reason.
\begin{itemize}
\item Is it at least 8 pages long? If not, exclude.
\item Is it published at one of the included venues (see Appendix~\ref{app:venues})? If not, exclude. Note that only the technical tracks at the conferences are included, no workshops or ‘special’ tracks (like SEET at ICSE). Only exclude based on the Publication Title field, or on the venue name printed in the PDF – no extra online search.
\item If there are no student subjects or no empirical study, exclude (this is also caught by the search later on).
\end{itemize}
\item Extract study information. First, do keyword search. If necessary (if any of the five categories below remain unanswered), read introduction, method (usually needed for the study type), and threats to validity.
\begin{itemize}
\item Which study type is it? Use the same term as used in the paper (e.g., ``Controlled experiment'', ``Family of quasi-experiments'', ``qualitative study''). If there is no empirical study conducted, exclude the paper.
\item How many students are participating in the study? If students and
professional subjects participate, count only the students. If it cannot clearly be determined how many students participated, use 'NA' (as long as it’s clear that students indeed participated, otherwise exclude). \newline
\textbf{Keyword search (one after the other, until you find the
information):} student, subject, graduate, master, bachelor, recruit, invit, participa
\item Was it voluntary to participate? Possible values: voluntary (also if it is mentioned that subjects were recruited openly, e.g., ``via flyers'',
``university-wide'', ``via mailing lists''), part of a course (if it is only mentioned that participation was in the scope of a course, not whether it was voluntary or mandatory), mandatory, and NA. \newline
\textbf{Keyword search:} Same as for voluntariness. Additionally: withdraw, option, volunt, mandatory, compulsory
\item How were (student) subjects compensated? Use the same term as used in the paper. If it cannot clearly be determined, use 'NA'.\newline
\textbf{Keyword search:} financ, money, monetary, reward, extra, bonus, compensat, voucher, receiv, gift, points
\item Was ethical approval obtained? Yes, No, Other, or NA. 'Other' if it is described that mandatory procedures (e.g., from university, state, or
country) were followed, but it is clear that this is not an ethical approval.\newline
\textbf{Keyword search:} ethic, irb, board, approv, consent
\end{itemize}
\end{enumerate}

\section{Questionnaire}
\label{app:questionnaire}
The online questionnaire consisted of the following questions. The entire questionnaire with introduction and final page, and the precise layout is found in the online dataset~\cite{datasetThis}.

\begin{enumerate}
\item In which country are you employed?\newline
If several, state the country in which you spend most of your work time. \newline\textbf{(Free-text with suggestions based on standard country list)}
\item What is your highest degree?\newline
\textbf{(Higher education entrance qualification, Bachelor degree (or equivalent), Master degree (or equivalent), PhD degree (or equivalent), Habilitation degree (or equivalent), Other (free text))}
\item What is your current, or last held, academic rank?\newline
(For example, doctoral student, associate professor)\newline
\textbf{(Free text)}
\item Have you previously conducted an empirical study which involved student subjects?\newline
Note: We understand ``subject'' broadly in this survey. That is, studies such as a case study involving students would also be included.\newline
\textbf{(Yes/No)}
\end{enumerate}

\noindent \textbf{New Page}

\begin{enumerate}
\setcounter{enumi}{4}
\item Which of the following statements regarding ethics approval processes applies to your work?\newline
\textbf{(Single selection with free-text option)}
\begin{itemize}
\item For studies with student subjects, I am required to obtain ethical approval
\item I am not required to obtain ethical approval, but there are mandatory steps for human-subject studies from my employer/the state (clarify details, if any)
\item Neither of the above (clarify details, if any)
\end{itemize}

\item How many studies with student subjects have you performed in the last 5 years?\newline
\textbf{($>$ 10 studies, 6-10 studies, 1-5 studies, None)}

\item In your most recent research study with students, did you recruit your own students?\newline
We refer to ``own'' students as students that are either supervised (thesis project, PhD supervision) by the respondent, or take a course
given by the respondent.\newline
\textbf{(Yes/No)}

\item In your most recent research study with students, was their participation mandatory or voluntary?\newline
\textbf{(Voluntary, Voluntary, but part of a graded course component (e.g., assignment), Mandatory)}

\item In your most recent research study with students, what forms of compensation did you offer to student subjects?\newline
Please mark all options that apply.\newline
\textbf{(None (besides potential learning experience), Bonus points in course, Monetary reward (e.g., fixed rate, a raffle/lottery), Snacks/Food, Other (please specify))}

\item In your most recent research study with students, did the subjects in your study/studies give consent to participate?\newline
\textbf{(Yes/No)}

\item In your most recent research study with students, did the subjects in your study/studies have the option to withdraw from the
study at any time?\newline
\textbf{(Yes/No)}

\item Based on your answers above, would you like to clarify any details?\newline
\textbf{(Free text)}
\end{enumerate}

\noindent \textbf{New Page}

\begin{enumerate}
\setcounter{enumi}{12}
\item Please state your agreement to the following statements regarding educational value of research studies.\newline
\textbf{(Per item: 5-point Likert agreement scale and ``don't know'' option)}
\begin{itemize}
\item If course content is prescribed by a curriculum, it should not be
changed just to make a research study fit in the course.
\item Research studies should be connected to course projects.
\item  Reviewers (of a study design/protocol) should pay close attention to students receiving adequate educational value from the research study.
\end{itemize}
\item Please state your agreement to the following statements regarding consent.\newline
\textbf{(Per item: 5-point Likert agreement scale and ``don't know'' option)}
\begin{itemize}
\item Every research study involving student subjects should be based on informed consent.
\item  Student subjects should be permitted to withdraw from a research study at any time.
\item Participation in a research study may be mandatory if it fits into the course context.
\item Participation in a research study conducted in a course should
always be voluntary for students.
\end{itemize}

\item Please state your agreement to the following statements regarding the course-study relationship.\newline
The following statements all apply to a situation in which a course instructor conducts a research study within his/her course.\newline
\textbf{(Per item: 5-point Likert agreement scale and ``don't know'' option)}\newline
From an ethical standpoint, it is acceptable...
\begin{itemize}
\item to use the enrolled students as subjects in a research study.
\item to base a part of the assessment on the participation in a research study.
\item to base a part of the assessment on the performance in a research study.
\item to encourage students to participate in a research study.
\item to withhold information from the students with respect to the study goals.
\item to withhold information from the students as to how the data they
provide will be used.
\item for the instructor to know who participated in the research study.
\item that the researcher conducting the research study is the same
person as the course instructor.
\end{itemize}

\item According to your opinion, how relevant is it to include the following information in a publication that uses student subjects?\newline
\textbf{(Per item: 5-point Likert scale from ``very
irrelevant'' to ``very relevant'', and ``don't know'' option)}
\begin{itemize}
\item Status of ethical approval or similar measures.
\item Voluntariness of student participation.
\item Compensation of student subjects.
\item Use of informed consent.
\end{itemize}

\item Do you have additional comments?\newline
\textbf{(Free text)}
\end{enumerate}

\section{Declarations}
\subsection{Funding}
University faculty funding with no external funding involved.

\subsection{Conflicts of interest/Competing interests}
Not applicable.

\subsection{Availability of data and material}
The data used in this manuscript is published on Zenodo \cite{datasetThis}.

\subsection{Code availability}
Not applicable.
\end{document}